%% file: main.tex
\documentclass[10pt,twocolumn,letterpaper]{article}

\usepackage[pagenumbers]{iccv} 


\definecolor{iccvblue}{rgb}{0.21,0.49,0.74}
\usepackage[pagebackref,breaklinks,colorlinks,allcolors=iccvblue]{hyperref}

\usepackage{multirow}
\usepackage{multicol}
\usepackage{algorithm}
\usepackage{algpseudocode}
\usepackage{amsmath}

\def\paperID{15741} 
\def\confName{ICCV}
\def\confYear{2025}

\title{ClimateGS: Real-Time Climate Simulation with 3D Gaussian Style Transfer}

\author{Yuezhen Xie\\
\\
\\
\and
Meiying Zhang\\
\\
\\
\and
Qi Hao\\
\\
\\
}

\begin{document}
\maketitle
\input{sec/00_abstract}
\input{sec/01_intro}
\input{sec/02_related}
\input{sec/03_preliminary}
\input{sec/04_method}
\input{sec/05_experiments}
\input{sec/06_conclusion}
{
    \small
    \bibliographystyle{ieeenat_fullname}
    \bibliography{main}
}

\clearpage
\input{sec/07_appendix}

\end{document}

%% file: sec/00_abstract.tex
\begin{abstract}

Adverse climate conditions pose significant challenges for autonomous systems, demanding reliable perception and decision-making across diverse environments. To better simulate these conditions, physically-based NeRF rendering methods have been explored for their ability to generate realistic scene representations. However, these methods suffer from slow rendering speeds and long preprocessing times, making them impractical for real-time testing and user interaction. This paper presents ClimateGS, a novel framework integrating 3D Gaussian representations with physical simulation to enable real-time climate effects rendering. The novelty of this work is threefold: 
1) developing a linear transformation for 3D Gaussian photorealistic style transfer, enabling direct modification of spherical harmonics across bands for efficient and consistent style adaptation; 
2) developing a joint training strategy for 3D style transfer, combining supervised and self-supervised learning to accelerate convergence while preserving original scene details; 
3) developing a real-time rendering method for climate simulation, integrating physics-based effects with 3D Gaussian to achieve efficient and realistic rendering. 
We evaluate ClimateGS on MipNeRF360 and Tanks and Temples, demonstrating real-time rendering with comparable or superior visual quality to SOTA 2D/3D methods, making it suitable for interactive applications.
\end{abstract}

%% file: sec/01_intro.tex
\section{Introduction}
\label{sec:intro}

Climate simulation plays a crucial role in the development and testing of autonomous systems, especially under extreme weather conditions, by enhancing system robustness through comprehensive evaluation of various environmental factors. Although significant progress has been made in NeRF-based climate simulation\cite{li2023climatenerf} by combining style transfer and physical modeling to improve realism, its long preprocessing and rendering times limit its feasibility in real-time applications.

\begin{figure}[t]
\centering
\includegraphics[width=0.99\linewidth]{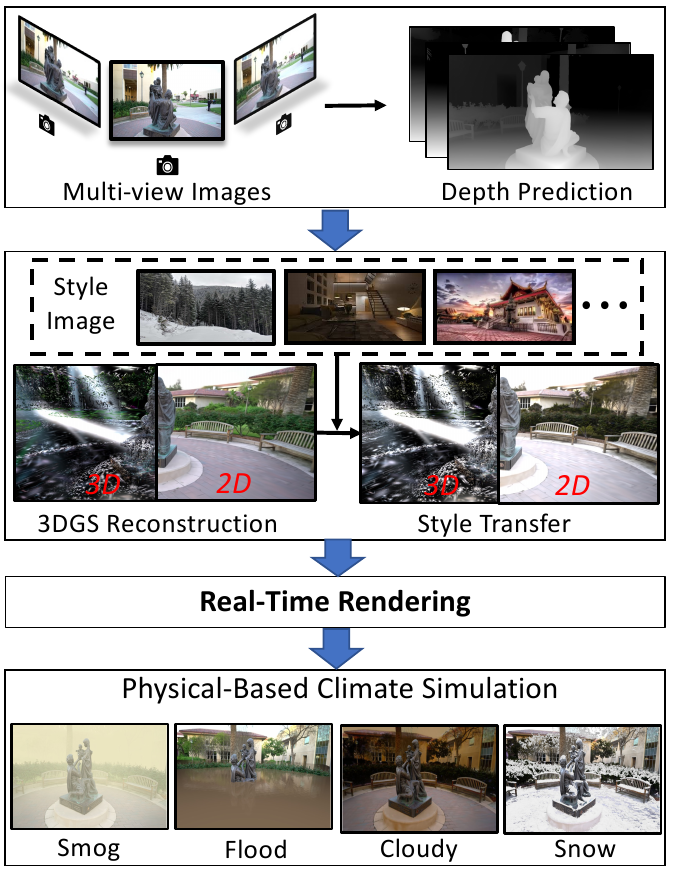}
{\begin{center}
\vspace{-0.3cm}
\caption{\textbf{Illustration of ClimateGS Pipeline.}
Our method consists of three stages: (a) Multi-view image-based reconstruction of the 3D scene; (b) Style transfer applied to the reconstructed scene for subsequent climate simulation (\cref{sec:style_transfer,sec:training_strategy}); (c) Physics-based climate simulation using a real-time renderer (\cref{sec:climate_simulation}). This process enables the generation of diverse climate simulations.
}
\label{fig:system}
\end{center}
}
\vspace{-1.0cm}
\end{figure}

To address these limitations, 3D Gaussian\cite{kerbl20233d} provide a more computationally efficient alternative. Building on this foundation, we propose a framework based on 3D Gaussians, as illustrated in \cref{fig:system}, which consists of three key components: (a) 3D Gaussian scene reconstruction with depth regularization, (b) 3D Gaussian-based style transfer, and (c) physics-based climate rendering. Specifically, we focus on optimizing three critical processes that impact both computational efficiency and rendering quality:
\begin{itemize}
\item \textbf{Limitations of Photorealistic 3D Scene Style Transfer:}
Existing 3D scene style transfer methods\cite{chen2024upst, wang2024gaussianeditor, zhang2023transforming, wu2024gaussctrl, li2023climatenerf} often rely on 2D image-based techniques, where colors are iteratively optimized while preserving scene geometry. These approaches are computationally expensive and often suffer from temporal inconsistencies, causing artifacts such as blurry textures and color shifts across views. While zero-shot 3D style transfer\cite{liu2024stylegaussian, kim2024fprf, zhang2023transforming, liu2023stylerf} avoid these problems, most approaches focus on artistic stylization rather than realistic effects, highlighting the need for photorealistic 3D style transfer.

\item \textbf{Lack of Detail Preservation in Style Transfer:}
A common approach integrates pre-trained VGG features into the radiance field\cite{kim2024fprf, miao2024conrf, liu2024stylegaussian} and applies 2D-based style transfer techniques such as AdaIN. However, this method lacks scalability and expressiveness, particularly in large-scale outdoor environments. Moreover, slight inconsistencies in multi-view images caused by variations in camera angles and environmental factors lead to discrepancies in VGG features across viewpoints. This inconsistency complicates gradient-based optimization, often resulting in color shifts and texture oversmoothing, which degrade structural integrity and fine-grained scene details.

\item \textbf{Inefficient Climate Simulation:}
Existing NeRF-based climate simulation methods\cite{li2023climatenerf} typically rely on computationally expensive volumetric rendering and physical-based models, such as metaballs for snow and FFT-based wave simulations for water effects. While these approaches produce highly realistic weather phenomena, their high computational cost makes them impractical for real-time applications. 
\end{itemize}

To address these challenges, ClimateGS introduces a novel photorealistic style transfer in 3D scenes and physics-driven rendering, enabling an interactive and real-time climate simulation. Our key contributions are as follows:

\begin{enumerate}
\item Developing a linear transformation for 3D Gaussian photorealistic style transfer, enabling direct modification of spherical harmonics across different bands for efficient and consistent style transfer while preserving view-dependent effects.
\item Developing a joint training strategy for photorealistic 3D style transfer, incorporating supervised learning for style consistency and self-supervised learning for detail preservation, to accelerate convergence while maintaining fine-grained scene details.
\item Developing a real-time climate simulation rendering method based on 3D Gaussian, integrating physically-based climate effects to achieve high-quality climate phenomena rendering.
\end{enumerate}

%% file: sec/02_related.tex
\section{Related Work}
\label{sec:related}

\begin{figure*}[t]
\centering
\includegraphics[width=0.99\linewidth]{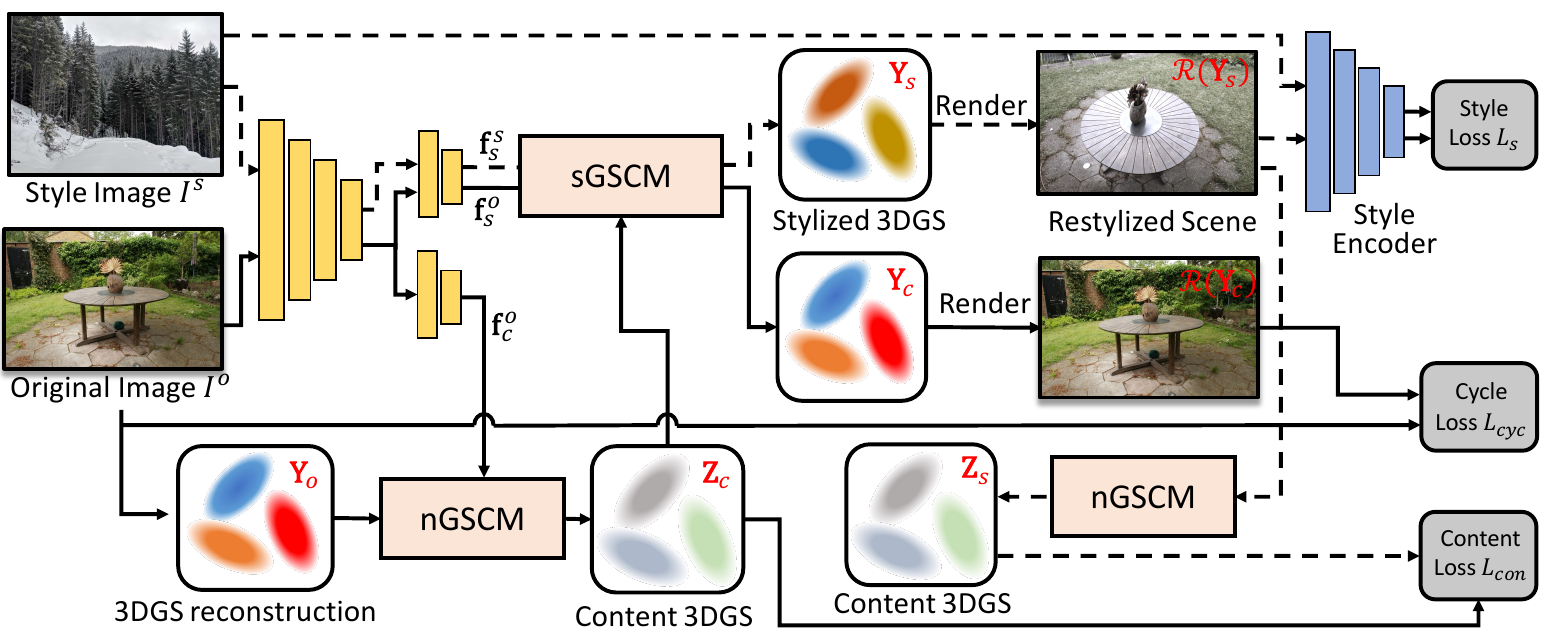}
{\begin{center}
\vspace{-0.3cm}
\caption{\textbf{Overview of Style Transfer Training Framework.}}
\label{fig:style_transfer_framework}
\end{center}
}
\vspace{-1.0cm}
\end{figure*}

{
\setlength{\parindent}{0cm}
\textbf{Photorealistic 3D Scene Style Transfer:}
Photorealistic 3D scene style transfer can be achieved through text-to-image methods\cite{wu2024gaussctrl, wang2024gaussianeditor, haque2023instruct, chen2024dge}. However, these methods typically rely on iterative optimization, which results in slower speeds. \cite{liu2023stylerf, huang2022stylizednerf} given a style image, can modify the scene's color to match the style in a feed-forward manner, \cite{liu2024stylegaussian} can operate at real-time speeds. However, this approach often focuses on artistic effects rather than photorealistic effects. Forward style transfer methods based on NeRF \cite{kim2024fprf, li2024instant} are constrained by the use of AdaIN, often resulting in blurry outputs with a slight bias toward artistic styles. 
To address these issues, we have developed a feed-forward photorealistic style transfer network based on 3D Gaussians, which can perform inference at real-time speeds.
}

{
\setlength{\parindent}{0cm}
\textbf{3D Scene Editing of Radiance Fields:}
3D scene editing in radiance fields has attracted considerable attention, with approaches incorporating physical properties \cite{verbin2022ref, huang2022hdr} or utilizing 2D generative models \cite{wang2022clip, haque2023instruct, liu2024genn2n}. However, NeRF-based methods are limited by implicit representations, resulting in high computational costs for optimization. The advent of 3D Gaussian representations \cite{kerbl20233d, yu2024mip, fan2025lightgaussian} mitigates these challenges, offering more efficient editing capabilities. Recent works \cite{chen2024dge, wang2024gaussianeditor}, such as GaussCtrl \cite{wu2024gaussctrl} optimize 3D Gaussians to alter scene attributes, while others \cite{zhou2024feature, guo2024semantic, ye2024gaussian} focus on tasks like object removal and semantic segmentation. In contrast, our method preserves the original scene structure, with an emphasis on integrating and rendering climate-related physical entities.
}

{
\setlength{\parindent}{0cm}
\textbf{Climate Simulation:}
Climate simulation in 2D images primarily relies on generative models. ClimateGAN\cite{schmidt2022climategan} employs depth prediction to generate water surface masks and integrates GANs for realistic water effects. \cite{li2021weather} utilizes CycleGAN\cite{zhu2017unpaired} for weather transformation based on diverse weather datasets. \cite{oh2024monowad, sakaridis2018semantic} generate foggy scenes using depth estimation. Additionally, BrushNet\cite{ju2024brushnet} can leverages text-to-image diffusion inpainting models for flood image generation. While these methods effectively simulate climate variations in single images, the absence of physical simulation limits their spatiotemporal consistency.
}

{
\setlength{\parindent}{0cm}
\textbf{Physical-based Climate Simulation:}
Many computer graphics studies focus on physically modeling climate phenomena. For instance, \cite{nishita1987shading} simulates uniform smog effects, and ClimateNeRF\cite{li2023climatenerf} builds on these methods for realistic rendering. Fourier transforms \cite{fournier1986simple} are used to simulate water surface ripples, while metaballs \cite{stomakhin2013material, nishita1997modeling} model snowfall. However, Fourier transforms are computationally expensive, and metaballs, when rendered independently, incur high costs due to their inability to integrate directly into the rendering pipeline. In contrast, we use Gerstner waves \cite{tessendorf2001simulating} for surface wave simulation, which is more computationally efficient and provides an effective numerical solution for surface normals. Additionally, we employ 3D Gaussians to simulate melting snow within metaballs, preserving snow properties. Our approach efficiently simulates physical climate effects in 3D Gaussian scenes while maintaining real-time rendering performance.
}

%% file: sec/03_preliminary.tex
\section{Preliminary: 3D Gaussian Splatting}
\label{sec:preliminary}

3D Gaussian Splatting\cite{kerbl20233d} represents a 3D scene as a set of 3D Gaussian primitives, where each Gaussian $G_j$ is described by a finite set of parameters: the center position $\mu_j$, covariance matrix $\Sigma_j$, opacity $\alpha_j$, and a view-dependent color function $sh_j$. To ensure the covariance matrix $\Sigma_j$ remains positive definite during the optimization process, it can be expressed as the product of an orthogonal rotation matrix $R$ and a diagonal scaling matrix $S$, that is $\Sigma = RSS^TR^T$.

Rendered through an optimized rasterization renderer, the color of a given pixel $p$ can be obtained by calculating the colors of the $N$ ordered Gaussians that cover the pixel from a specific viewpoint:
\begin{equation}
  c(p) = \sum_{j=1}^N sh_j(\theta, \phi) \alpha_j \prod_{k=1}^{j-1}(1-\alpha_k)
  \label{eq:render_gs}
\end{equation}

The contribution of each Gaussian to the color of pixel $p$ depends on the viewing direction $(\theta, \phi)$. The color function $c_j(\theta, \phi)$ associated with the Gaussian depends on the spherical coordinates, and is given by an $l$-degree spherical harmonic function:
\begin{equation}
  c_j(\theta, \phi) = sh_{j, 0} + \sum^3_{l=1}\sum_{m=1}^{2l+1}sh_{j,l,m}Y_{l,m}(\theta, \phi)
  \label{eq:render_sh}
\end{equation}
where the vectors $ sh_{j,0}, sh_{j,l,m} \in \mathbb{R}^3 $ denote the color of each band, and $ Y_{l,m} $ form a basis of the spherical harmonics polynomials of degree $l$. In short, $ sh_{j,0} \in \mathbb{R}^3 $ represents the primary color, while the additional coefficients $ sh_{j,l,m}$ encode the smooth variation of the color as the viewing direction changes.

%% file: sec/04_method.tex
\section{Proposed Method}
\label{sec:method}


\subsection{Photorealistic 3D Gaussian Style Transfer}\label{sec:style_transfer}

The color tones of images captured on snowy days differ significantly from those taken in spring or summer. This discrepancy cannot be effectively addressed through physical modeling, making style transfer a more suitable solution. However, existing methods for photorealistic style transfer have yet to be adapted for 3D Gaussians. To bridge this gap, we are inspired from recent advancements in 2D photorealistic scene style transfer \cite{ke2023neural, li2019learning} and propose a zero-shot photorealistic style transfer method for 3D Gaussians, illustrated in \cref{fig:style_transfer_framework} . Our approach ensures real-time transfer efficiency while minimizing memory consumption, effectively preserving color fidelity and preventing blurring.

\subsubsection{Color Space Transformation}\label{sec:color_trans}
Unlike previous style transfer pipelines, which embed 2D image features directly into 3D Gaussians, our approach formulates style transfer as a linear mapping function $\mathbf{F}$. This function applies a purely linear transformation to the color of each 3D Gaussian. Specifically, for a given 3D Gaussian $G_j$, its color $c_j(\theta, \phi)$ is represented by the spherical harmonic coefficients $sh_{j,l,m}$, allowing the style transfer process to be expressed as:
\begin{equation}
  \mathbf{F}(c_j) = \mathbf{F}(sh_{j,0}) + \sum^3_{l=1}\sum_{m=1}^{2l+1}\mathbf{F}(sh_{j,l,m})Y_{l,m}
  \label{eq:linear_transfer}
\end{equation}

If all spherical harmonic coefficients can be mapped into a common color space, the style transfer process can be formulated as a single linear transformation $\mathbf{F}$. Since the basis transformations $Y_{l,m}$ induced by viewpoint changes only scale the color values, explicit handling of view variations is unnecessary. Therefore, if all SH coefficients $sh_j$ can be directly mapped to the RGB color space $sh_j' \in \mathbb{R}^{(2l+1) \times 3}$, then a uniform transformation can be applied across different basis coefficients:
\begin{equation}
  sh_j' = \Lambda \cdot sh_j
  \label{eq:sh_transfer}
\end{equation}
Here, $\Lambda \in \mathbb{R}^{(2l+1)}$ represents predefined spherical harmonic polynomials, serving as a color space transformation. Thus, the same linear transformation $\mathbf{F}$ can be consistently applied to SH coefficients across different bases.

\subsubsection{3D Gaussian Color Mapping (GSCM)}
We utilize the network architecture proposed in \cite{ke2023neural} due to its efficiency and streamlined design. Furthermore, we introduce an enhanced version of the Deterministic Neural Color Mapping module, referred to as the 3D Gaussian Color Mapping (GSCM) module.

\begin{figure}[t]
\centering
\includegraphics[width=0.99\linewidth]{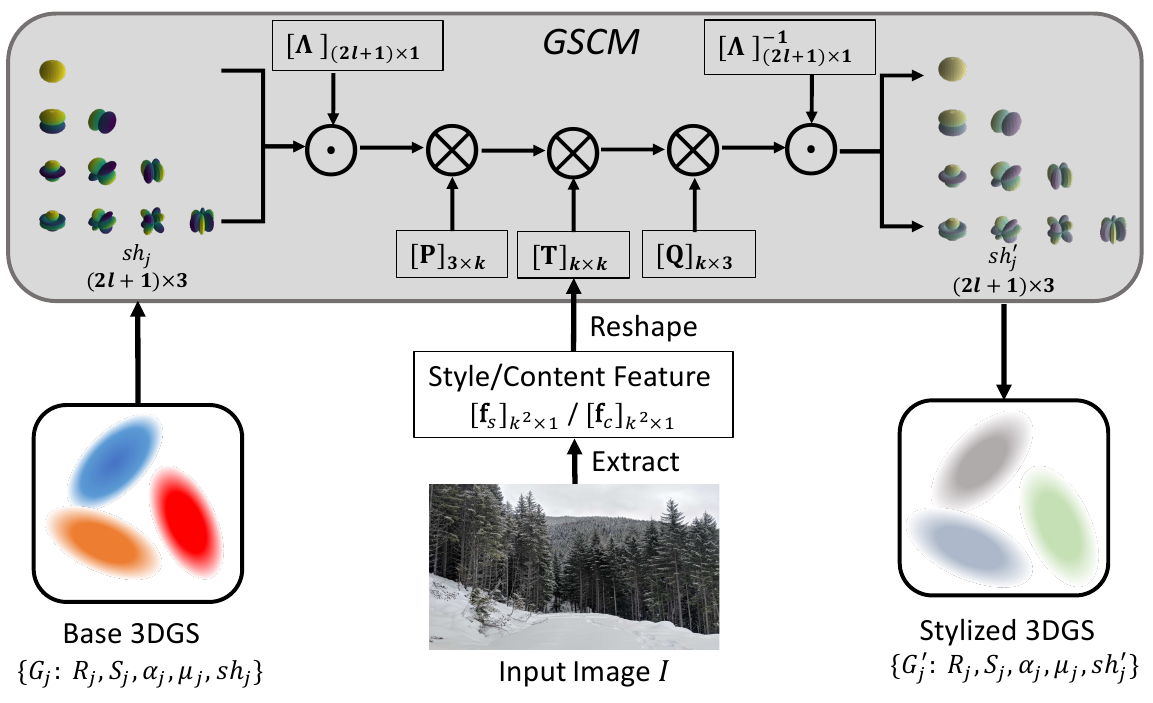}
{\begin{center}
\vspace{-0.3cm}
\caption{\textbf{Illustration of GSCM.} 
The GSCM module consists of three color projection matrices ($\Lambda$, $\mathbf{P}$, $\mathbf{Q}$) and a transformation matrix $\mathbf{T}$, mapping 3D Gaussian color $sh_j$ from a unified space to the color space defined by the style image, and then restore it to the spherical harmonic space using the inverse of matrix $\Lambda$.
}
\label{fig:GSCM}
\end{center}
}
\vspace{-1.0cm}
\end{figure}

As \cref{fig:GSCM} illustrated, given a reconstructed Gaussian set $G_j$ and an input style image $\mathbf{I}$, we first feed $\mathbf{I}$ into an encoder $\mathbf{E}$, which predicts an image-adaptive color transformation matrix $\mathbf{T}$ of size $(k \times k)$ and reshapes it into $(k, k)$:
\begin{equation}
  \mathbf{E}(\mathbf{I}) = \mathbf{T}^{(k \times k)} \rightarrow \mathbf{T}^{(k, k)} 
  \label{eq:style_encoder}
\end{equation}

For each input Gaussian $G_j$, we only focus on transforming its color $sh_j$. The spherical harmonic coefficients $sh_j$ are initially mapped to a unified space $sh_j'$ using $\Lambda$ as mentioned in \cref{sec:color_trans}, then projected into a $k$-dimensional space via $\mathbf{P}^{(3,k)}$. The coefficients are transformed by $\mathbf{T}^{(k,k)}$, projected back to RGB space through $\mathbf{Q}^{(k,3)}$, and finally restored to the spherical harmonic domain using $\Lambda^{-1}$:
\begin{equation}
    \text{GSCM}(G_j, \mathbf{T}) = sh_j \cdot \Lambda^{(2l+1,)} \cdot \mathbf{P}^{(3, k)} \cdot \mathbf{T}^{(k, k)} \cdot \mathbf{Q}^{(k, 3)} \cdot \Lambda^{-1}
  \label{eq:style_DNCM}
\end{equation}

We also employ nGSCM to map the color of the Gaussians to the content space, and sGSCM to transform the content into the target color style space. The image $\mathbf{I}$ is passed through the encoder $\mathbf{E}$, which generates two color transformation matrices ${\mathbf{f}_c, \mathbf{f}_s}$, corresponding to nGSCM and sGSCM, respectively. The final transferred Gaussian $\mathbf{Y}$ is then given by:
\begin{equation}
    \mathbf{Y} = \mathbf{F}(c_j) = \text{sGSCM}(\text{nGSCM}(\mathbf{Y}_o, \mathbf{f}_c^o), \mathbf{f}_s^s)
  \label{eq:style_sDNCM}
\end{equation}
Here, $\mathbf{E}(\mathbf{I}^o)$ outputs ${\mathbf{f}_c^o, \mathbf{f}_s^o}$, where $\mathbf{I}^o$ represents the content image, analogous to the style image $\mathbf{I}^s$, and $\mathbf{Y}_o$ denotes the reconstructed Gaussian model.

\subsection{Joint Training Strategy}\label{sec:training_strategy}

As shown in \cref{fig:style_transfer_framework}, the first stage of style transfer normalizes the color of each Gaussian to ensure the transferred Gaussians $\mathbf{Y}_s$ stay consistent with $\mathbf{Y}_o$ in content space. When mapping $\mathbf{Y}_s$ back to the content space $\mathbf{Z}_s$, their values should match the content of reconstructed Gaussians $\mathbf{Z}_c$. Thus, we apply an L1 loss between these values. Unlike 2D style transfer, 3D Gaussians can render images, so the rendered images $\mathcal{R}(\mathbf{Z}_s)$ and $\mathcal{R}(\mathbf{Z}_c)$ should remain consistent, facilitating faster convergence:
\begin{equation}
    \mathcal{L}_{con} = ||\mathbf{Z}_c - \mathbf{Z}_s||_1 + ||\mathcal{R}(\mathbf{Z}_c) - \mathcal{R}(\mathbf{Z}_s)||_1
  \label{eq:consistency_loss}
\end{equation}
where $\mathcal{R}(\cdot)$ denotes the rendering process. The style loss is formulated as:
\begin{equation}
    \begin{aligned}
      \mathcal{L}_s = \sum^n_{i=1}||\mu(\phi_i(\mathcal{R}(\mathbf{Y}_s))) -\mu(\phi_i(I^s))|| + \\
      \sum^n_{i=1}||\sigma(\phi_i(\mathcal{R}(\mathbf{Y}_s))) - \sigma(\phi_i(I^s))||
    \end{aligned}
  \label{eq:style_loss}
\end{equation}
where $I^s$ denotes style image.

Since all modules are interconnected, the original Gaussian $\mathbf{Y}_o$ can be reconstructed by first normalizing the Gaussian using $\mathbf{f}_c^o$, then applying nGSCM with $\mathbf{f}_s^o$ to recolor it and get $\mathbf{Y}_c$, similar to whitening, and recoloring process in WCT\cite{li2018closed}. To ensure a more stable initial training process and accelerate network convergence, we introduce cycle consistency loss:
\begin{equation}
    \mathcal{L}_{cyc} = ||\mathbf{Y}_c - \mathbf{Y}_o||_1 + ||\mathcal{R}(\mathbf{Y}_c) - I^o||_1
  \label{eq:cycle_loss}
\end{equation}
The total loss is a combination of three losses:
\begin{equation}
  \mathcal{L} = \mathcal{L}_s + \lambda_{con}\mathcal{L}_{con} + \lambda_{cyc}\mathcal{L}_{cyc}
  \label{eq:total_loss}
\end{equation}
where $\lambda_{con}, \lambda_{cyc}$ are the loss weights.

\subsection{Climate Effect Simulation}\label{sec:climate_simulation}
In a reconstructed 3D Gaussian scene, rendering techniques are employed to simulate climate effects. During forward pass, we obtain $\mathbf{\alpha}, \mathbf{c}, \mathbf{d}$ for a given pixel $p$, representing its transmittance, color, and depth, respectively. The rendered depth is computed using the method outlined in \cite{kerbl2024hierarchical}:
\begin{equation}
    \mathbf{d} = \sum^N_{i=1}T_i\alpha_id_i, 
    \;\; \text{where} \;\;
    d_i=(R_ip_i + t_i)_z.
  \label{eq:render_depth}
\end{equation}

Hence, in subsequent climate simulations, deferred rendering can be utilized to efficiently apply different climate effects in real time. Compared to similar approaches in NeRF \cite{li2023climatenerf}, 3D Gaussian-based deferred rendering can significantly accelerate the rendering process.

\begin{algorithm}[t]
\caption{Gumbel Distribution Algorithm}
\label{alg:gumbel}
\begin{algorithmic}[1]

\Require \\
  $D$: 3D Gaussians depth sequence;\\
  $N$: Number of 3D Gaussians;\\
  $\alpha$: Light transmittance of Gaussians;
\Ensure
  optimal $d$

\State initial $d=0$, $W=0$ and $T=1$;
\State initial $\mu=0$, $\beta=0.2$ and $w=0$;
\State $Q \sim \mathbf{Gumbel}(\mu, \beta)$

\For{$t = 1,\dots,N$}
    \State $x \gets D_t$

    \State Compute PDF: $ P(x | Q) = \frac{1}{\beta} e^{-(x - \mu)/\beta} e^{-e^{-(x - \mu)/\beta}}$
    \If{$P(x | Q) < 0.5$}
        \State Update $d \gets \mu$, $W \gets w$ if $W \leq w$
        \State Reset $\mu = 0$ and $\beta = 0.2$;
    \Else
        \State $w \gets w + \alpha_t \cdot T$
        \State Incrementally update $\mu$ and $\beta$ using $x$ and $\alpha_t * T$
    \EndIf
    \State $T \gets T \cdot (1 - \alpha_t)$
\EndFor
\end{algorithmic}
\end{algorithm}

\subsubsection{Smog Simulation}
We model smog effects with uniformly distributed microscopic particles, assuming zero radiance contribution from 3D Gaussians in the region before depth $\mathbf{d}$. To enhance efficiency while preserving absorption characteristics, we approximate the radiative transfer equation (RTE) \cite{lenoble1985radiative} by neglecting scattering and applying the Beer-Lambert law \cite{beer1852bestimmung}, yielding:
\vspace{-0.2cm}
\begin{equation}
    F(p) = (1-w)\mathbf{c}(p) + w \cdot c_\text{smog}, 
    \;\; \text{where} \;\;
    w=e^{-\sigma_\text{smog}\mathbf{d}}.
\label{eq:fog_simulation}
\end{equation}
where $c_\text{smog} \in \mathbb{R}^3$ denotes smog color, $\sigma_\text{smog} \in \mathbb{R}^+$ controls smog density. Both parameters can be controlled by users.

\subsubsection{Flood Simulation}
Following \cite{li2023climatenerf}, the water surface is defined by a central point $\mathbf{o}_w$ and a normal vector $\mathbf{n}_w$, approximated by the plane equation $\mathbf{n}_w \cdot (x - \mathbf{o}_w) = 0$.

To simulate waves efficiently, we adopt the Gerstner wave model\cite{tessendorf2001simulating}, which computes spatiotemporal surface normals from wave direction, steepness, and wavelength. Multiple waves are combined for realistic undulations, with an analytical solution for normals that avoids numerical differentiation, reducing computational cost.

For reflection and refraction, we apply the Fresnel equations using the Schlick approximation\cite{schlick1994inexpensive}, sampling both reflected and refracted rays. Mipmap-accelerated Screen Space Reflections further enhance rendering efficiency.

\subsubsection{Snow Simulation}
We model snow accumulation using 3D Gaussians, leveraging their blending properties to balance realism and efficiency. Traditional metaball-based volumetric rendering is computationally expensive, making real-time performance infeasible.

Snow Gaussians are placed via the ray-splat intersection method mentioned in \cite{gao2024relightable} with parallel projection. To mitigate floating artifacts caused by inaccurate depth estimation, we model depth distributions with multiple Gumbel distributions, selecting the mode of the highest-transmittance Gumbel as the final depth. As \cref{alg:gumbel} illustrates, an incremental computation scheme further optimizes efficiency by initializing new Gumbel distributions only when depth deviations exceed a predefined threshold.

\begin{figure}[t]
\centering
\includegraphics[width=0.99\linewidth]{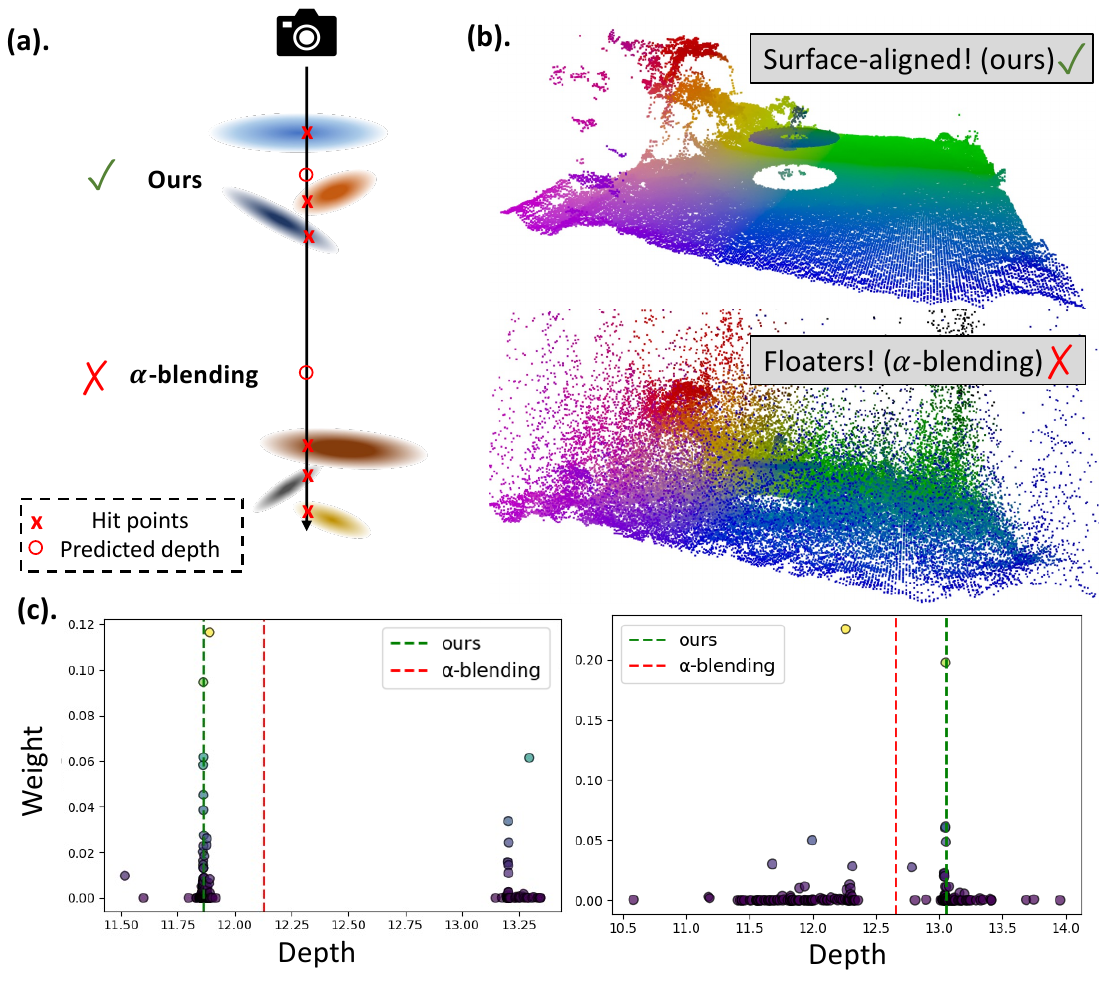}
{\begin{center}
\vspace{-0.7cm}
\caption{\textbf{Procedure of Snow Sampling.}
(a) demonstrates our method produces more accurate results compared to $\alpha$-blending; (b) and (c) show that on real-world data, our approach reduces erroneous floaters and provides more accurate depth estimation.
}
\label{fig:snow_sampling}
\end{center}
}
\vspace{-1.0cm}
\end{figure}

\cref{fig:snow_sampling} shows that this approach substantially improves depth estimation accuracy, ensuring precise placement of snow 3D Gaussians while reducing floating artifacts.

In a fixed viewpoint, the color of a single 3D Gaussian remains constant, making it unsuitable for simulating subsurface scattering. Without subsurface scattering, snow appears overly white and unrealistic. To approximate this effect with minimal computational overhead, we propose a method utilizing normal estimation and wrap lighting\cite{green2004real} for subsurface scattering simulation.

For normal computation within a 3D Gaussian, the normal at the Gaussian center, when projected onto the image plane, must be aligned with the ray direction connecting this point to the camera. To facilitate computation, all subsequent calculations are conducted in view space.

\begin{figure}[t]
\centering
\includegraphics[width=0.99\linewidth]{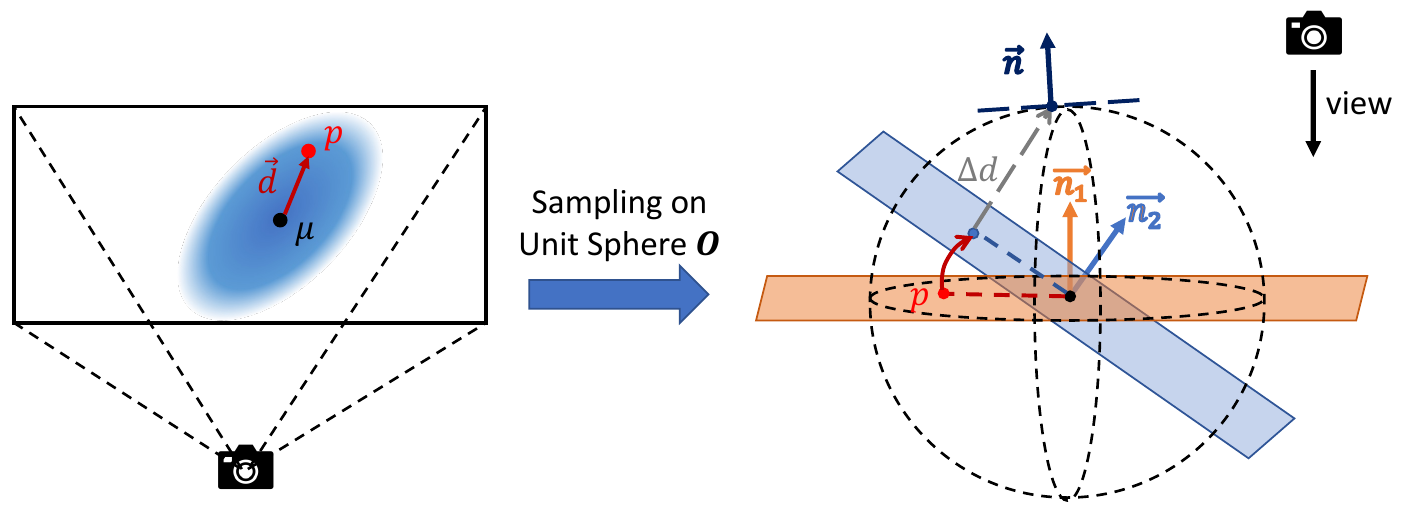}
{\begin{center}
\vspace{-0.3cm}
\caption{\textbf{Gaussian Normal Sampling Process.}}
\label{fig:normal_sampling}
\end{center}
}
\vspace{-1.0cm}
\end{figure}

To estimate the relative position of a pixel $p$ with respect to the 3D Gaussian center $\mu$, we define the Mahalanobis distance $\mathbf{D}$ as:
\vspace{-0.2cm}
\begin{equation}
\mathbf{D} = \sqrt{(p-\mu)^T\Sigma^{-1}(p-\mu)}
\label{eq:mahalanobis_dist}
\vspace{-0.2cm}
\end{equation}
where $\mathbf{D}$ provides a normalized measure of the pixel's distance from the 3D Gaussian center. Let $\vec{d}$ denotes the vector from the 3D Gaussian center to the projected point $p$. Given the hit point direction $\vec{d}$ and its normalized distance $\mathbf{D}$, the corresponding point can be mapped onto the unit sphere \( \mathbf{O} \), from which the normal is subsequently sampled.

As illustrated in \cref{fig:normal_sampling}, to align the normal of the projected point with the ray direction from the 3D Gaussian center to the camera, we project $\vec{d}$ onto the plane defined by the Gaussian center and the camera-ray direction:
\vspace{-0.2cm}
\begin{equation}
  \vec{d'} = \vec{d} - \frac{\vec{d} \cdot \vec{n_2}}{\vec{n_1} \cdot \vec{n_2}}
  \label{eq:proj_vec}
  \vspace{-0.2cm}
\end{equation}
where $\vec{n_1} = (0,0,-1)$ is a unit vector along the negative $\mathbf{Z}$-axis, and $\vec{n_2}$ is the vector from the 3D Gaussian center to the camera. The length of the edge parallel to $\vec{n_2}$ is given by $\Delta d = \sqrt{1 - \mathbf{D}^2}$.

Using this, we can directly compute points on $\mathbf{O}$ and subsequently obtain the sampled normal:  
\vspace{-0.2cm}
\begin{equation}
  \vec{n} = \mathbf{D} \cdot \vec{d'} + \Delta d \cdot \vec{n_2}
  \label{eq:normal_sample}
  \vspace{-0.2cm}
\end{equation}

Finally, the normal is transformed into world space using a rotation matrix $R\vec{n}$. This approach efficiently approximates subsurface scattering by treating the 3D Gaussian as a sphere, sampling normals, and applying wrap lighting, significantly improving snow realism.

%% file: sec/05_experiments.tex
\section{Experiments}
\label{sec:experiments}

This section evaluates photorealistic style transfer and climate simulation. For style transfer, we compare with SOTA 3D methods, achieving higher quality and faster inference. For climate simulation, we benchmark against SOTA 2D synthesis and 3D editing methods, demonstrating superior rendering speed, quality, and consistency. Results show that ClimateGS achieves comparable or superior performance in physical realism, spatiotemporal coherence, and computational efficiency.

\subsection{Experiment Setup}
{
\setlength{\parindent}{0cm}
\textbf{Datasets.}
We evaluate photorealistic style transfer on two real-world datasets, Tanks and Temples\cite{Knapitsch2017} and MipNeRF360\cite{barron2022mipnerf360}, using MS COCO\cite{lin2014microsoft} as the style image. Since ClimateGS is designed for outdoor scenes, we testing on Playground, Family, Horse, and Train from Tanks and Temples, Garden from MipNeRF360, and Seq-0031 from the Waymo Open dataset\cite{Sun_2020_CVPR}.
}


For scene reconstruction prior to climate simulation, NeRF-based methods use Instant-NGP\cite{muller2022instant}, while 3D Gaussian methods (including ours) use 3D Gaussian Splatting\cite{kerbl20233d} with depth priors. Depth regularization is applied using different supervision sources: Depth-Anything-V2\cite{yang2024depth} for MipNeRF360 and Tanks and Temples, and Street Gaussians\cite{yan2024street} with LiDAR supervision for Waymo.

{
\setlength{\parindent}{0cm}
\textbf{Baselines.}
For photorealistic style transfer, we compare our method with two SOTA zero-shot radiance field approaches: \textbf{FPRF} \cite{kim2024fprf}, which embeds VGG features into K-Planes\cite{fridovich2023k}, and \textbf{StyleGaussian} \cite{liu2024stylegaussian}, which integrates downsampled VGG features into 3D Gaussians. As no existing method directly applies zero-shot photorealistic style transfer to 3D Gaussians, StyleGaussian serves as the primary baseline. Optimization-based methods \cite{chen2024upst, mei2025regs, wu2024gaussctrl} are excluded due to high computational costs.
}

For climate effect simulation, we compare ClimateGS with SOTA 2D editing methods (\textbf{BrushNet} \cite{ju2024brushnet}, \textbf{ClimateGAN} \cite{schmidt2022climategan}) and 3D scene editing methods (\textbf{ClimateNeRF} \cite{li2023climatenerf}, \textbf{GaussCtrl} \cite{wu2024gaussctrl}). BrushNet utilizes latent diffusion for image inpainting, ClimateNeRF simulates climate via phically-based, and GaussCtrl performs text-guided 3D scene editing (e.g., "make it snowed") by ControlNet\cite{zhang2023adding}.

\begin{figure*}[t]
\centering
\includegraphics[width=0.99\linewidth]{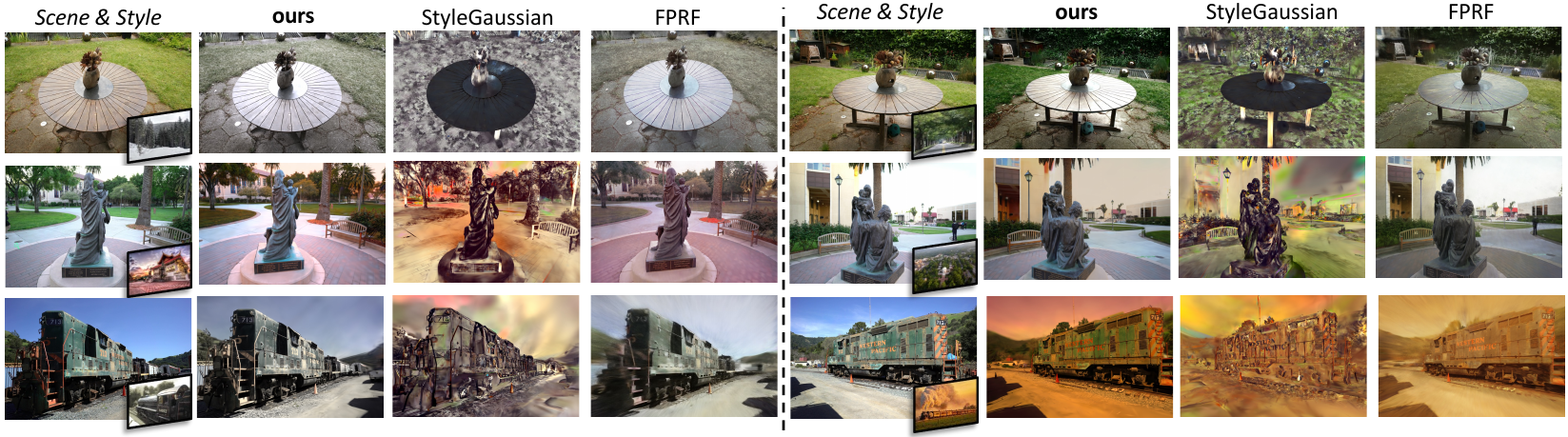}
{\begin{center}
\vspace{-0.5cm}
\caption{\textbf{Qualitative Comparsion.}
We compare our method with two zero-shot radiance field methods, StyleGaussian\cite{liu2024stylegaussian} and FPRF\cite{kim2024fprf}. Our method demonstrates superior style transfer quality, achieving better alignment with the style reference image while preserving the original textures and maintaining the inherent colors of the objects.
}
\label{fig:style_compare}
\end{center}
}
\vspace{-1.0cm}
\end{figure*}

{
\setlength{\parindent}{0cm}
\textbf{Implementation.}
Consistent with \textbf{NeuralPreset}\cite{ke2023neural}, we use EfficientNet-B0\cite{tan2019efficientnet} as the image encoder, with $k$ set to 16 and output dimension to 256. We train for 300,000 steps using the Adam optimizer\cite{kingma2014adam}, with an initial learning rate of $1e^{-4}$, decreasing by 0.1 every 100,000 steps. To ensure stability, we start training with only the cycle consistency loss $\mathcal{L}_{cyc}$, introducing content loss $\mathcal{L}_{con}$ and style loss $\mathcal{L}_s$ after 30,000 steps. We set $\lambda_{cyc}$ and $\lambda_{con}$ to 10.
}
\subsection{Style Transfer Results}

\begin{table}[t]
    \centering
    
    \resizebox{\columnwidth}{!}{
    
    \begin{tabular}{@{\hspace{0.7em}}c@{\hspace{0.7em}}|@{\hspace{0.7em}}c@{\hspace{0.7em}}c@{\hspace{0.7em}}c@{\hspace{0.7em}}c@{\hspace{0.7em}}|@{\hspace{0.7em}}c@{\hspace{0.7em}}c@{\hspace{0.7em}}c@{\hspace{0.7em}}}
    \toprule
    
    Methods & 
    \multicolumn{2}{c}{\begin{tabular}{@{}c@{}}Short-range\\Consistency\end{tabular}} & 
    \multicolumn{2}{c|@{\hspace{0.7em}}}{\begin{tabular}{@{}c@{}}Long-range\\Consistency\end{tabular}} & 
    \begin{tabular}{@{}c@{}}Transfer\\Time\end{tabular} & 
    \begin{tabular}{@{}c@{}}Rendering\\Time\end{tabular} &
    \begin{tabular}{@{}c@{}}Memory\\Usage\end{tabular} \\
    
    \midrule
    
    {}  & \textit{LPIPS} & \textit{RMSE}  & \textit{LPIPS} & \textit{RMSE} & \textit{Seconds} & \textit{Seconds} & \textit{GB}\\
    
    Original & 0.070 & 0.069 & 0.213 & 0.198 & - & 0.002 & 2.08 \\
    FPRF \cite{kim2024fprf} & 0.082 & 0.084 & \textbf{0.154} & \underline{0.165} & 0.129 & 18.851 & 12.79 \\
    StyleGaussian \cite{liu2024stylegaussian} & \underline{0.062} & \underline{0.057} & 0.181 & \textbf{0.164} & \underline{0.090} & \textbf{0.001} & \underline{8.30} \\
    \textbf{ours} & \textbf{0.057} & \textbf{0.053} & \underline{0.171} & 0.175 & \textbf{0.023} & \underline{0.002} & \textbf{3.24} \\
    \bottomrule
    \end{tabular}
    }

    \caption{\textbf{Quantitative results.}
    We evaluate the consistency performance of our method against state-of-the-art methods using LPIPS ($\downarrow$) and RMSE ($\downarrow$), and further assess its performance in transfer time, rendering time and memory usage.}
    
    \label{tab:style_comparsion}
\vspace{-0.7cm}
\end{table}

{
\setlength{\parindent}{0cm}
\textbf{Qualitative Results.}
\cref{fig:style_compare} illustrates the advantages of our method in style transfer. StyleGaussian fails to preserve color details, overly smoothing high-frequency textures and reducing scene realism. While FPRF uses AdaIN and VGG embedding-decoding, it inevitably introduces blurring. In contrast, our method better preserves scene details and intrinsic colors, ensuring high-quality style transfer. This is primarily achieved by formulating style transfer as a linear transformation, enabling consistent application across all spherical harmonic bands and avoiding color degradation caused by VGG decoding.
}

{
\setlength{\parindent}{0cm}
\textbf{Quantitative Results.}
As discussed in \cite{liu2024stylegaussian, kim2024fprf, chen2024upst}, no standardized metrics exist for evaluating 3D scene style transfer quality. Therefore, we conduct a quantitative analysis of multi-view consistency, transfer time, rendering time, and memory usage in \cref{tab:style_comparsion}. To ensure fairness, multi-view consistency is evaluated only for first-order spherical harmonics. Experimental results show that our method achieves comparable or superior consistency to SOTA. FPRF exhibits high long-term consistency, likely due to blurring introduced during the style transfer process. By directly storing transferred colors in 3D Gaussians, our method ensures strict cross-view consistency and significantly reduces transfer time and memory usage while maintaining rendering efficiency. In contrast, FPRF requires storing both VGG and semantic features, and StyleGaussian embeds VGG features within each 3D Gaussian, resulting in higher memory and time costs. Eliminating the need for pre-stored VGG features enhances the scalability of our method, making it more suitable for large-scale scenes.
}

{
\setlength{\parindent}{0cm}
\textbf{Ablation Study.}
The supplementary material provides a detailed analysis of the effectiveness of $\mathcal{L}_{con}$, $\mathcal{L}_{cyc}$, and the GSCM module. Please refer to it for further details.
}

\subsection{Climate Simulation Results}

\begin{figure}[t]
\vspace{-0.3cm}
\centering
\includegraphics[width=0.99\linewidth]{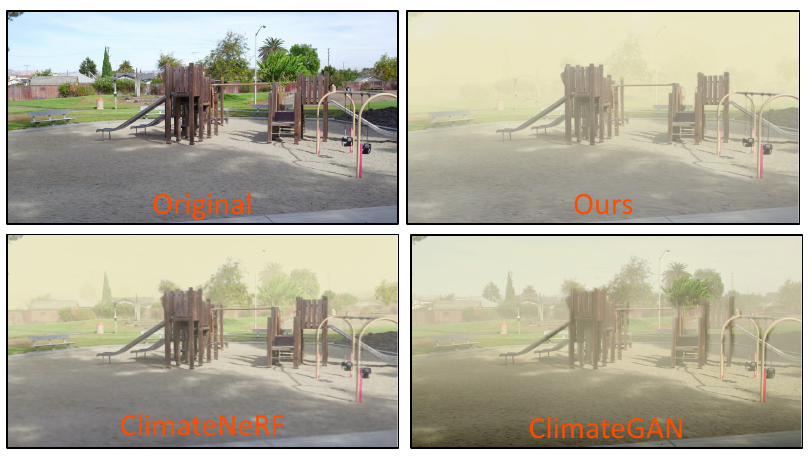}
{\begin{center}
\vspace{-0.3cm}
\caption{\textbf{Smog simulation comparison.}
ClimateGS simulates smog with view consistency, preserving clearer boundaries and more accurate object geometry.
}
\label{fig:smog_playground}
\end{center}
}
\vspace{-1.2cm}
\end{figure}

\begin{figure*}[t]
\centering
\includegraphics[width=0.99\linewidth]{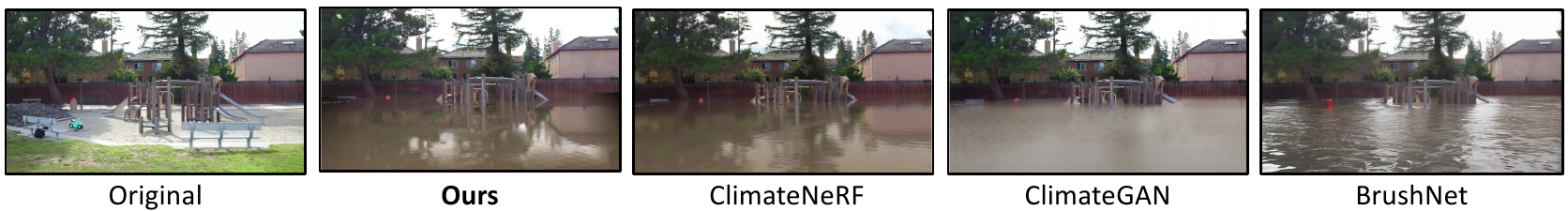}
{\begin{center}
\vspace{-0.5cm}
\caption{\textbf{Flood simulation comparison.} 
BrushNet and ClimateGAN generate realistic water surfaces, but are affected by random content and view inconsistencies. ClimateGS simulates realistic water ripples and reflections, achieving results comparable to ClimateNeRF.
}
\label{fig:flood_horse}
\end{center}
}
\vspace{-1.0cm}
\end{figure*}

\begin{figure}[h]
\centering
\includegraphics[width=0.9\linewidth]{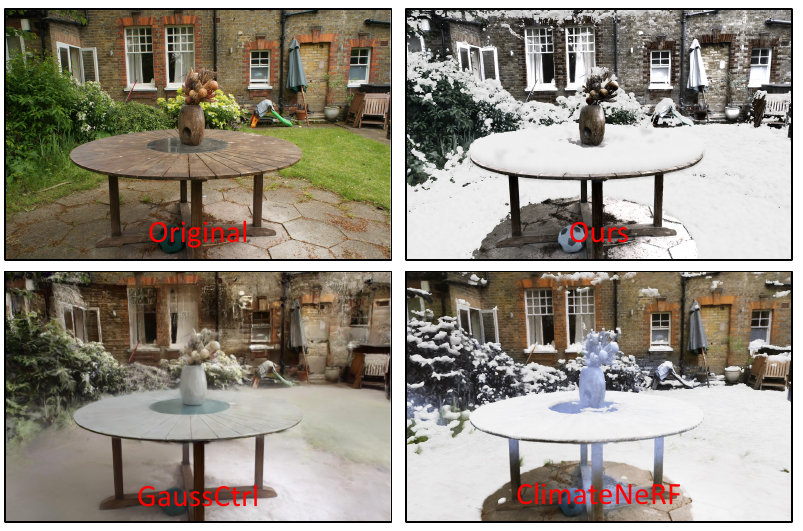}
{\begin{center}
\vspace{-0.3cm}
\caption{\textbf{Snow simulation comparison.}
ClimateGS effectively generates physical entities, correctly covers surfaces, and preserves a realistic winter tone.
}
\label{fig:snow_garden}
\end{center}
}
\vspace{-0.7cm}
\end{figure}

{
\setlength{\parindent}{0cm}
\textbf{Qualitative Results}
\cref{fig:smog_playground} presents the qualitative results of smog simulation, where ClimateGS and ClimateNeRF produce comparable outcomes. Due to depth map inaccuracies, ClimateGAN fails to generate sharp boundaries and temporally consistent results, leading to noticeable artifacts. 
}

\cref{fig:flood_horse} shows flood simulation results, where ClimateGAN and BrushNet produce inaccurate water reflections and lack frame-to-frame consistency, with BrushNet also introducing extraneous objects. ClimateGS accurately simulates water reflections, refractions, and ripples, enhancing realism, and achieves results comparable to ClimateNeRF with significantly lower resource consumption.

\cref{fig:snow_garden} shows snow simulation results. GaussCtrl merely modifies texture colors without incorporating snow as a physical entity, causing severe scene distortion. ClimateNeRF’s style transfer misapplies changes to artificial flowers on the table, resulting in unrealistic outputs. In contrast, ClimateGS adjusts the scenes tone and employs subsurface scattering to capture snow’s natural irregularities, producing a more convincing winter scene.

\begin{table}[t]
    \centering
    
\begin{tabular}{l|ccc}
\toprule
Method & Smog & Flood & Snow\\
\midrule
ClimateNeRF \cite{li2023climatenerf} & 1.687s & 12.062s & 1.5h / 4.374s\\
\textbf{ours} & \textbf{0.020s} & \textbf{0.020s} & \textbf{1m / 0.025s}\\
\bottomrule

\end{tabular}

\caption{\textbf{Quantitative results.} 
We evaluate the time required for simulation: smog and flood columns represent rendering time, the snow column lists preprocessing/rendering times separately.
}

\label{tab:climate_comparsion}
\vspace{-0.5cm}
\end{table}

{
\setlength{\parindent}{0cm}
\textbf{Quantitative Results}
\cref{tab:climate_comparsion} compares preprocessing and rendering times, showing that our method outperforms ClimateNeRF across three climate simulations. This improvement is attributed to the efficient rendering of 3D Gaussians and our optimizations for climate effect, enabling real-time performance. Meanwhile, a user study further validates our method’s effectiveness, with details provided in the supplementary materials.
}

\begin{figure}[h]
\centering
\vspace{-0.2cm}
\includegraphics[width=0.99\linewidth]{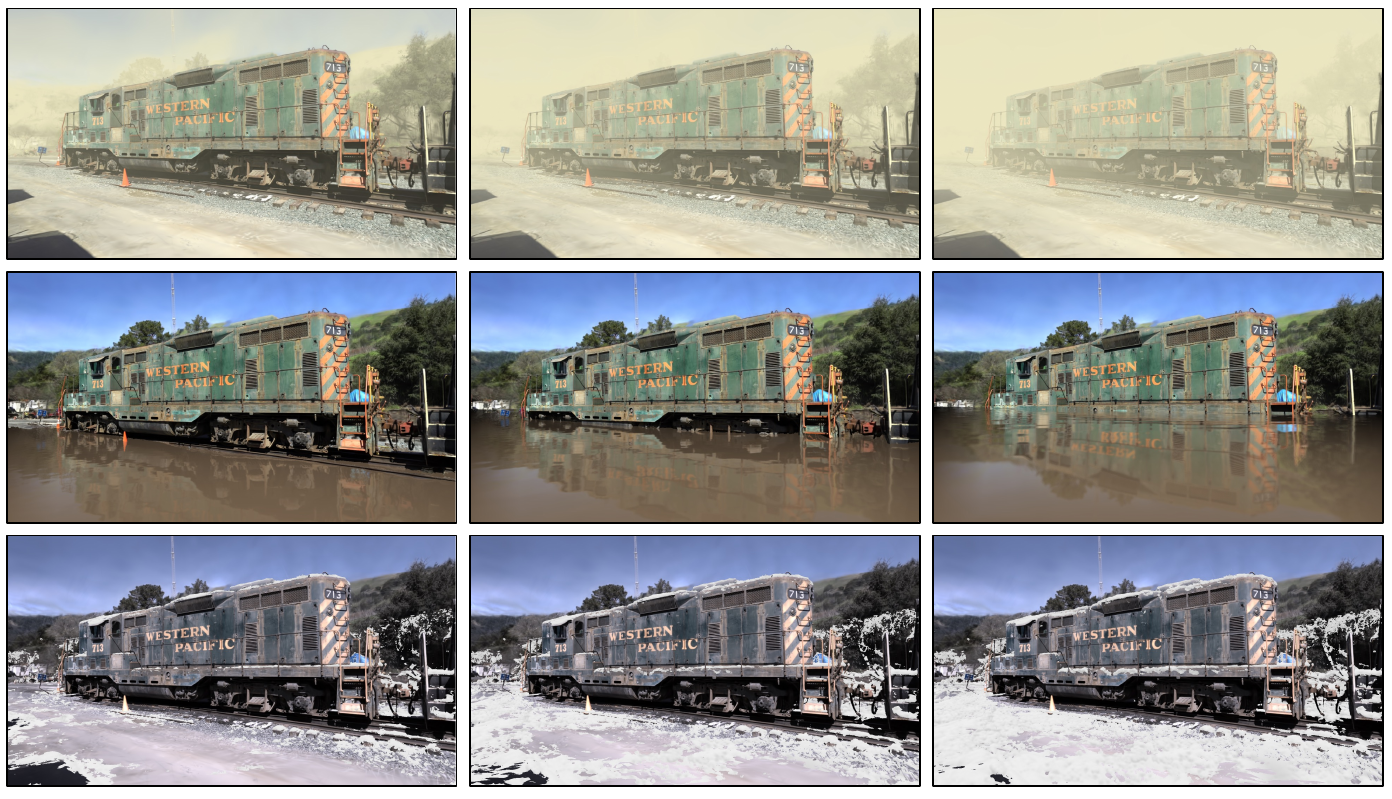}
{\begin{center}
\vspace{-0.3cm}
\caption{\textbf{Controllable Simulation.}
ClimateGS allows user-controlled real-time adjustments, including \textbf{smog density} (1st row), \textbf{flood levels} (2nd row), and \textbf{snow accumulations} (3rd row).
}
\label{fig:controllable}
\end{center}
}
\vspace{-0.7cm}
\end{figure}

{
\setlength{\parindent}{0cm}
\textbf{Controllability.}
ClimateGS enables real-time control over climate simulation effects, with \cref{fig:controllable} showing the impact of smog density, water level, and snow thickness. Compared to ClimateNeRF, ClimateGS supports real-time adjustments, making it ideal for autonomous driving simulations and interactive environments.
}

\begin{figure}[t]
\centering
\includegraphics[width=0.99\linewidth]{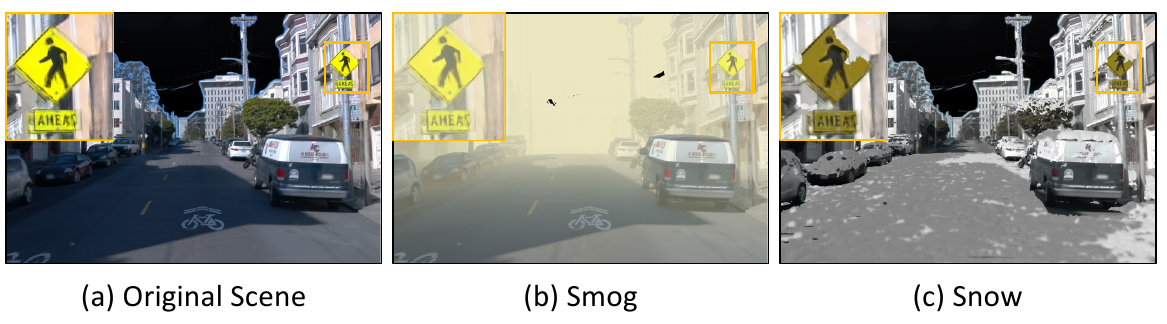}
{\begin{center}
\vspace{-0.3cm}
\caption{\textbf{Simulation on Driving Scenes\cite{Sun_2020_CVPR}.}}
\label{fig:autonomous_driving}
\end{center}
}
\vspace{-1.2cm}
\end{figure}

{
\setlength{\parindent}{0cm}
\textbf{Adverse weather simulation for self-driving.}
ClimateGS can be applied to autonomous driving for robust perception under multi-weather simulation. As shown in \cref{fig:autonomous_driving}, smog and snow degrade traffic sign recognition in Street Gaussians-reconstructed scenes, while training perception models in controlled simulations improves robustness in extreme weather. Notably, its separate skybox rendering pipeline results in missing sky regions.
}

{
\setlength{\parindent}{0cm}
\textbf{Limitations.}
In flood simulation, screen-space reflections cause inaccuracies in off-screen color rendering. In snow simulation, surface translucency affects 3D Gaussian visibility, limiting clear snow boundaries and snow details. Enhancing simulation accuracy while ensuring real-time performance remains a key research direction.
}

%% file: sec/06_conclusion.tex
\section{Conclusion}
\label{sec:conclusion}
This paper presents ClimateGS, a framework for climate simulation in 3D Gaussian scenes. It reformulates 3D Gaussian style transfer as a linear transformation for efficient and consistent style transfer while incorporating a joint training strategy to accelerate convergence and preserve scene details. Additionally, an optimized rendering pipeline enhances both efficiency and realism. Experiments on MipNeRF360 and Tanks and Temples datasets show that ClimateGS achieves comparable or superior visual quality to SOTA 2D/3D methods while delivering a 100× rendering speedup. Evaluations on the Waymo dataset further demonstrate its applicability to autonomous driving climate simulation. Future work will improve visual fidelity, such as realistic sky reflections in floods and more explicit snow boundaries, and enhance adaptability across diverse environments.

%% file: sec/07_appendix.tex
\appendix
\begin{center}
    \textbf{\Large Supplementary Material}
\end{center}
\section{Ablation Studys}
\begin{figure}[h]
\centering
\includegraphics[width=0.99\linewidth]{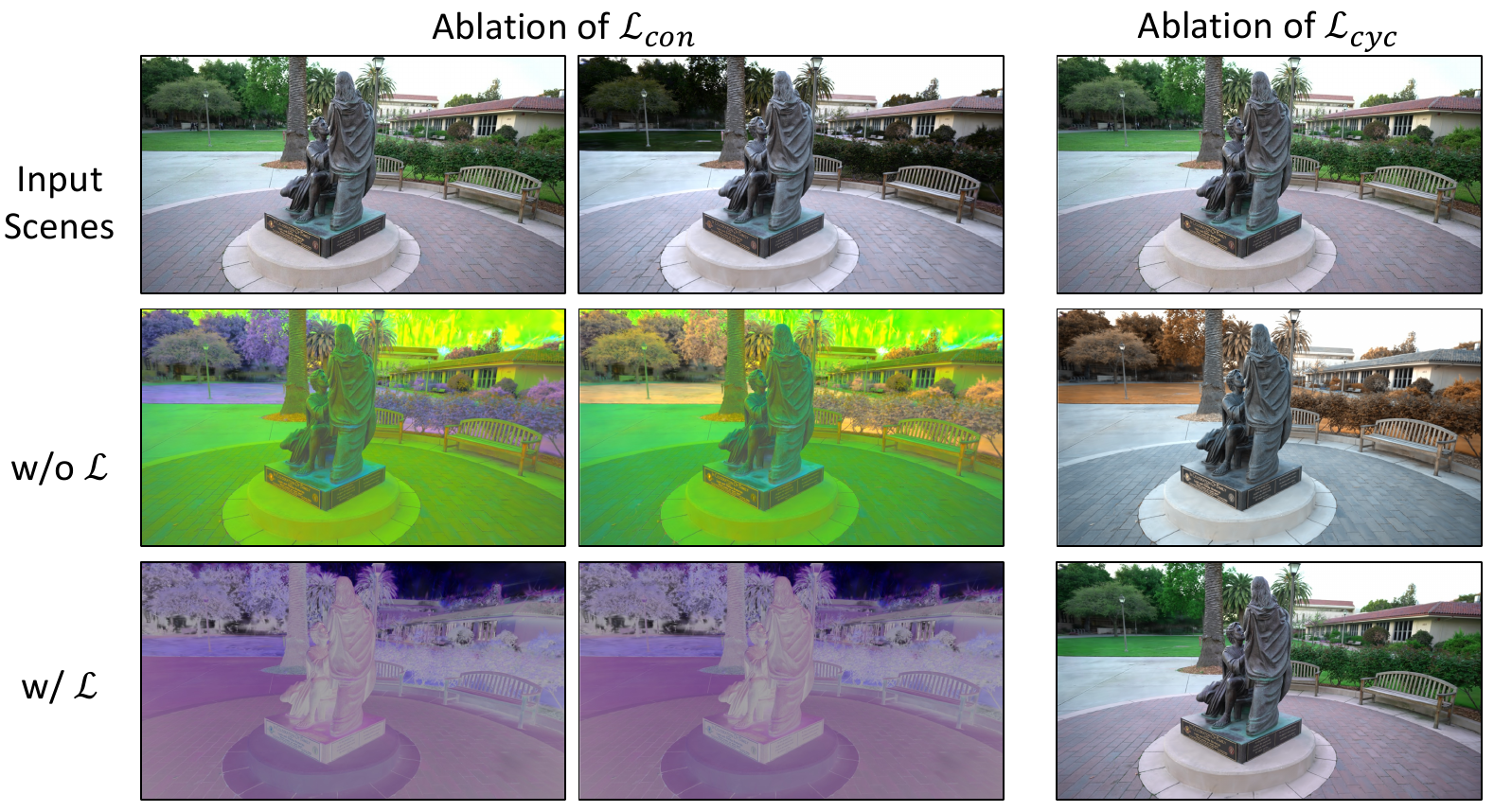}
{\begin{center}
\vspace{-0.5cm}
\caption{\textbf{Impact of $\mathcal{L}_{con}$ and $\mathcal{L}_{cyc}$.}
We demonstrate the colors of the original and transferred scenes mapped to the content space, where $\mathcal{L}_{con}$ ensures more consistent results. Right column shows the scenes mapped back to the original color space, where $\mathcal{L}_{cyc}$ mitigates errors introduced by the transformation.
}
\label{fig:ablation}
\end{center}
}
\vspace{-1.0cm}
\end{figure}

\paragraph{(1). Effect of $\mathcal{L}_{con}$ and $\mathcal{L}_{cyc}$.}
In \cref{fig:ablation}, we illustrate the impact of different losses. The left column shows the "scene content" from the first stage, where $\mathcal{L}_{con}$ ensures content independence from the transferred colors. The right column presents the second stage, where applying the transformation matrix and $\mathcal{L}_{cyc}$ restores the original colors, enhancing training stability and accelerating convergence.

\begin{figure}[h]
\centering
\includegraphics[width=0.99\linewidth]{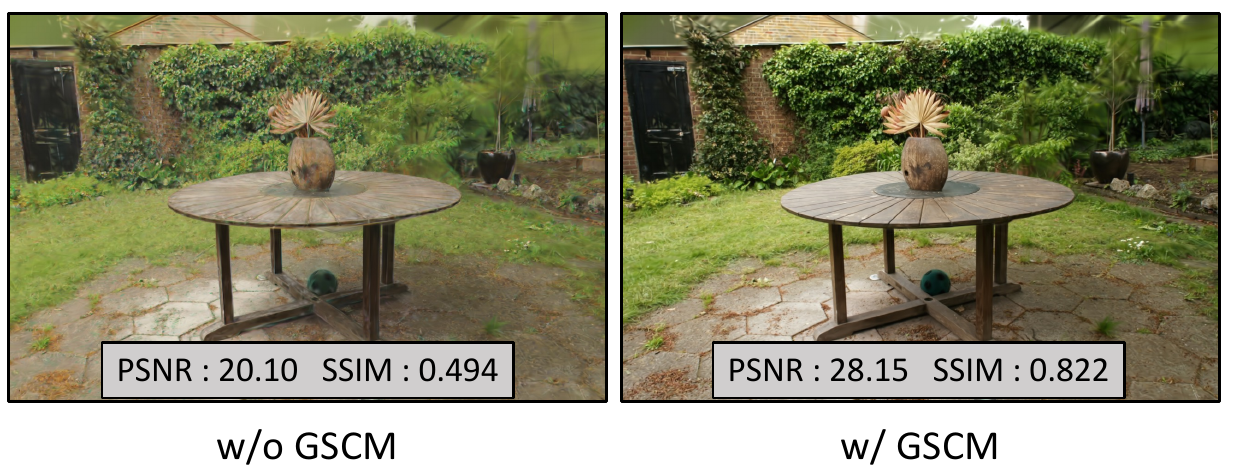}
{\begin{center}
\vspace{-0.3cm}
\caption{\textbf{Impact of GSCM.}
We demonstrate the effectiveness of the GSCM module, where its decoded colors outperform VGG-based methods in PSNR and SSIM metrics, better preserving the original scene details.
}
\label{fig:ablation_GSCM}
\end{center}
}
\vspace{-1.0cm}
\end{figure}

\paragraph{(2). Effect of GSCM.}
\cref{fig:ablation_GSCM} highlights the effectiveness of the GSCM module. Compared to VGG feature embedding in 3D Gaussian with MLP decoding as in \cite{kim2024fprf}, GSCM better preserves details and reduces blurriness, outperforming VGG-based decoding.

\subsection{User Study.}

\begin{figure}[h]
\centering
\includegraphics[width=0.99\linewidth]{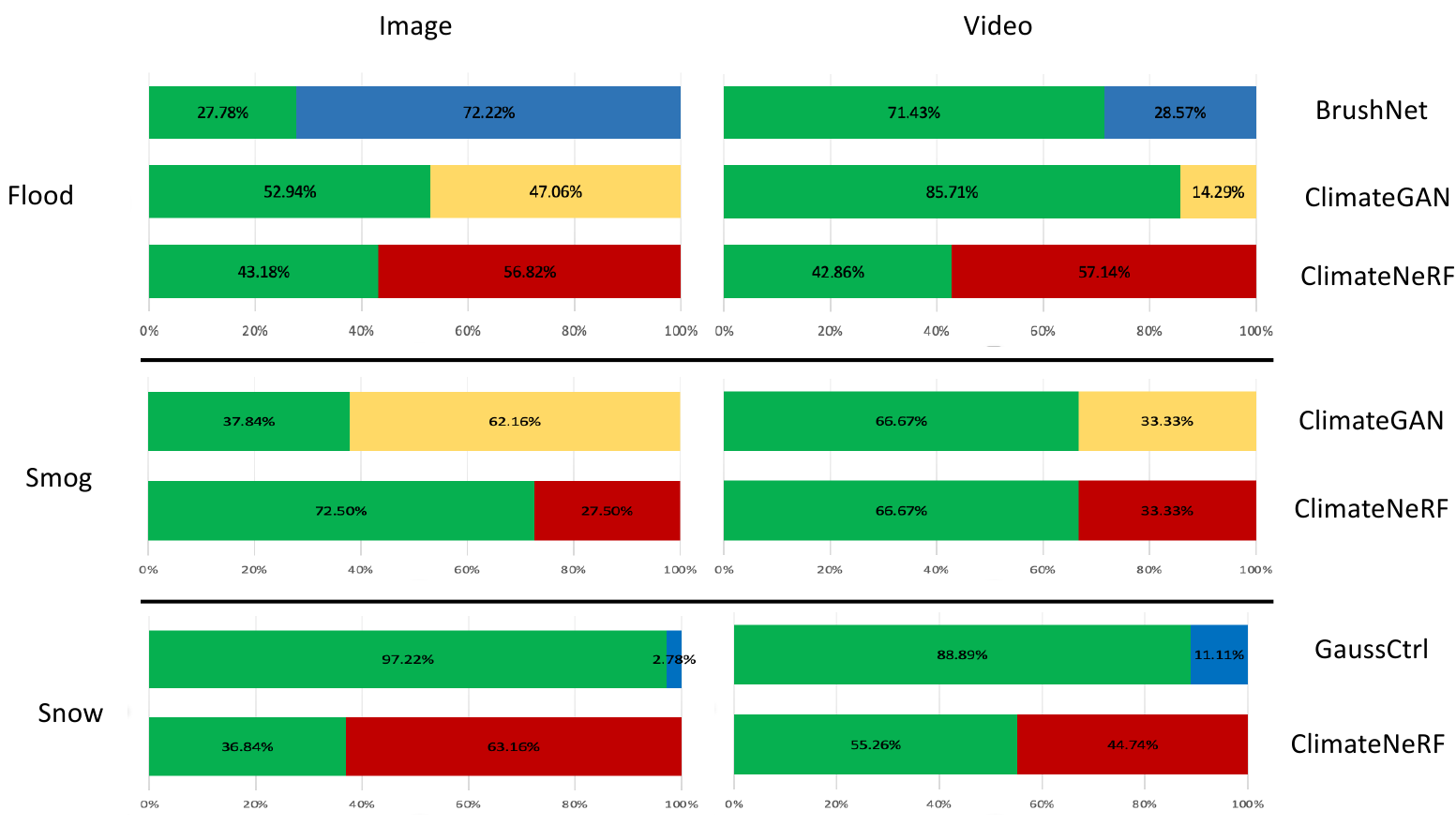}
{\begin{center}
\vspace{-0.3cm}
\caption{\textbf{User Study.}}
\label{fig:user_study}
\end{center}
}
\vspace{-1.0cm}
\end{figure}

We conducted a user study to quantitatively evaluate the realism of our method. Participants viewed paired synthetic images or videos of the same scene and selected the one that appeared more realistic. The study included three randomly selected images and a full video per scene, with a total of 17 participants. Results are shown in \cref{fig:user_study}. ClimateGS was preferred in most video comparisons due to its high realism and view consistency. While ClimateNeRF enhanced realism in flood simulations by reflecting off-screen objects, its blurry modeling reduced overall authenticity. In single-image comparisons, ClimateGS outperformed most baselines, while BrushNet produced more realistic ripples in flood scenes. However, generation-based methods fail to ensure multi-view consistency. ClimateGAN suffers from flickering, while BrushNet introduces additional objects, leading to significant inconsistencies.

\subsection{More Results.}
\begin{figure*}[t]
\centering
\includegraphics[width=0.99\linewidth]{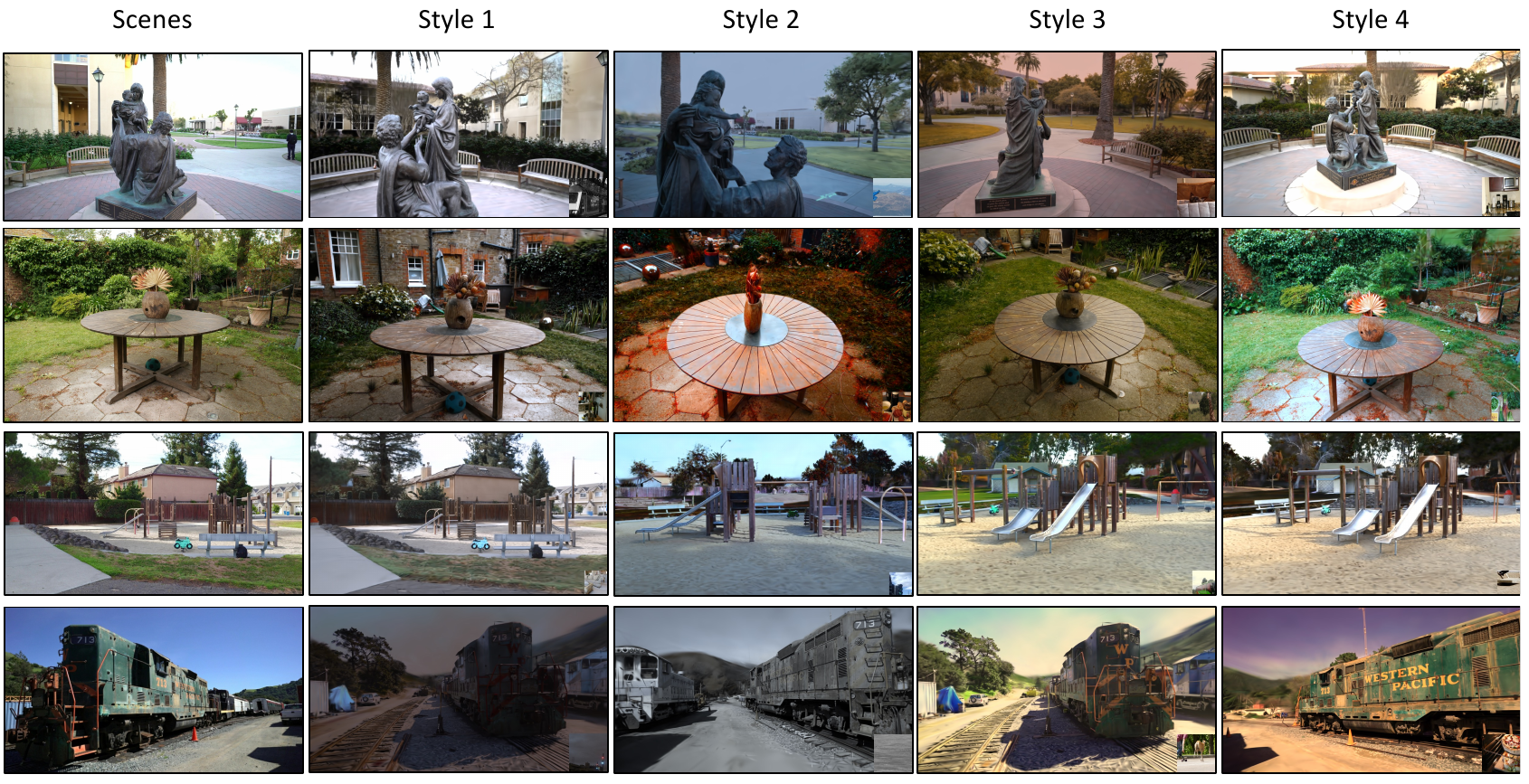}
{\begin{center}
\vspace{-0.3cm}
\caption{\textbf{Our Scene Transfer Results.}
Our proposed method effectively and stably transfers style colors across various scenes. The first column presents the original scenes to be transformed, while the subsequent columns display different style variations. The style reference image is shown in the bottom right corner.
}
\label{fig:appendix_style}
\end{center}
}
\vspace{-1.0cm}
\end{figure*}

\paragraph{Style Transfer}
In \cref{fig:appendix_style}, we present additional experimental results to demonstrate the effectiveness of our proposed style transfer method. As shown in the figure, our approach not only preserves fine details of the scene during style transfer but also maintains the original colors of objects while adapting to the target style. This advantage largely stems from our formulation of style transfer as a linear transformation, allowing each frequency band of the spherical harmonics to be modified in a consistent manner. As a result, our method avoids the color distortions typically introduced by VGG-based decoding.

\begin{figure*}[t]
\centering
\includegraphics[width=0.99\linewidth]{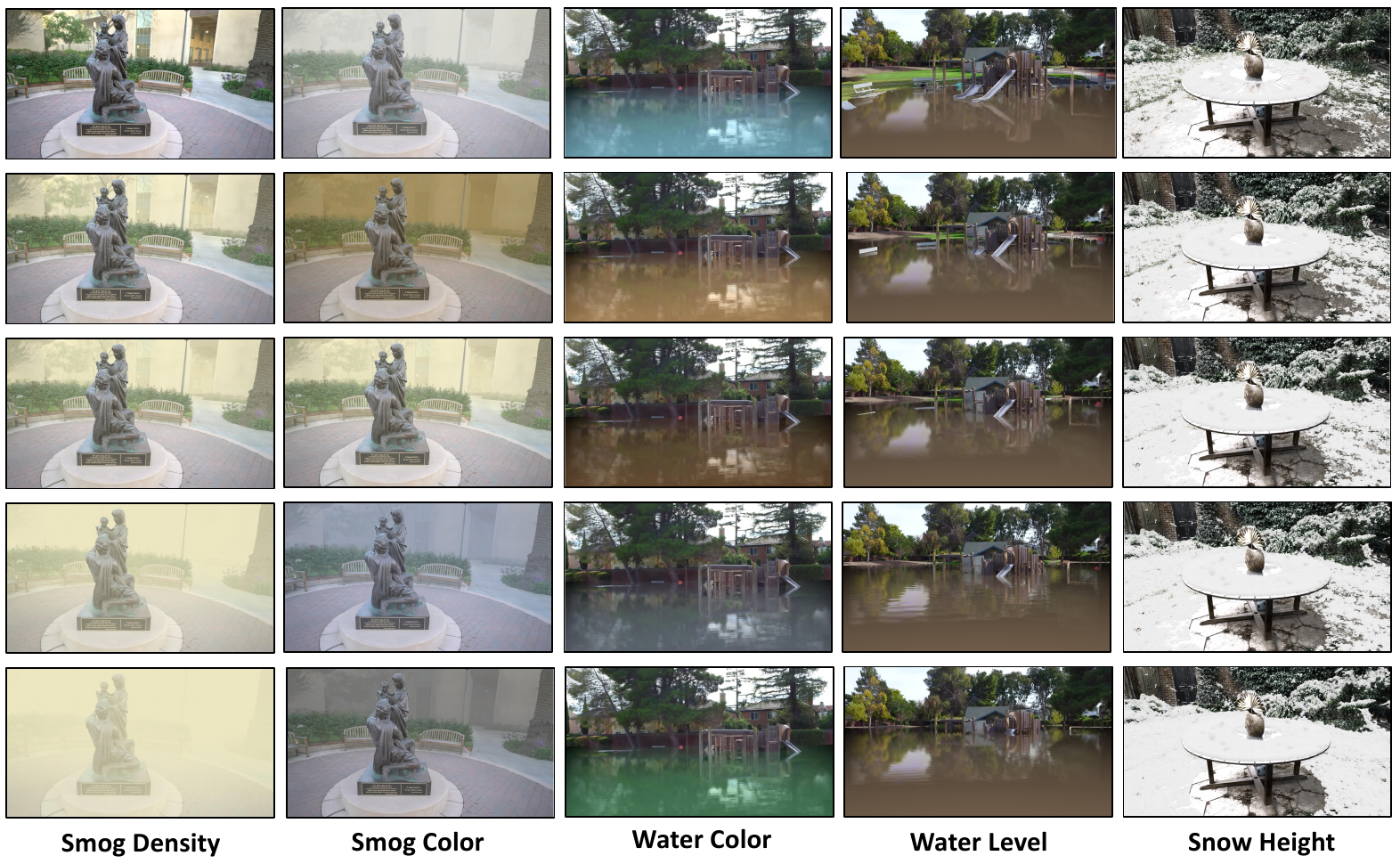}
{\begin{center}
\vspace{-0.3cm}
\caption{\textbf{Controllable Climate Simulation.}}
\label{fig:appendix_control}
\end{center}
}
\vspace{-1.0cm}
\end{figure*}

\paragraph{Climate Simulation}
We further demonstrate the high controllability of ClimateGS in climate simulation. In \cref{fig:appendix_control}, our method successfully simulates various environmental conditions, including different colors and densities of smog, varying water levels and colors of floods, and different depths of accumulated snow. The results indicate that our simulation framework allows for precise and user-controllable adjustments while enabling real-time modifications. Compared to ClimateNeRF, ClimateGS facilitates efficient real-time weather adaptation, making it particularly suitable for applications such as autonomous driving simulations and interactive virtual environments, thereby enhancing the overall user experience.

%% file: main.bbl
\begin{thebibliography}{55}
\providecommand{\natexlab}[1]{#1}
\providecommand{\url}[1]{\texttt{#1}}
\expandafter\ifx\csname urlstyle\endcsname\relax
  \providecommand{\doi}[1]{doi: #1}\else
  \providecommand{\doi}{doi: \begingroup \urlstyle{rm}\Url}\fi

\bibitem[Barron et~al.(2022)Barron, Mildenhall, Verbin, Srinivasan, and Hedman]{barron2022mipnerf360}
Jonathan~T. Barron, Ben Mildenhall, Dor Verbin, Pratul~P. Srinivasan, and Peter Hedman.
\newblock Mip-nerf 360: Unbounded anti-aliased neural radiance fields.
\newblock \emph{CVPR}, 2022.

\bibitem[Beer(1852)]{beer1852bestimmung}
Beer.
\newblock Bestimmung der absorption des rothen lichts in farbigen fl{\"u}ssigkeiten.
\newblock \emph{Annalen der Physik}, 162\penalty0 (5):\penalty0 78--88, 1852.

\bibitem[Chen et~al.(2024{\natexlab{a}})Chen, Laina, and Vedaldi]{chen2024dge}
Minghao Chen, Iro Laina, and Andrea Vedaldi.
\newblock Dge: Direct gaussian 3d editing by consistent multi-view editing.
\newblock In \emph{European Conference on Computer Vision}, pages 74--92. Springer, 2024{\natexlab{a}}.

\bibitem[Chen et~al.(2024{\natexlab{b}})Chen, Yuan, Li, Liu, Wang, Xie, Wen, and Yu]{chen2024upst}
Yaosen Chen, Qi Yuan, Zhiqiang Li, Yuegen Liu, Wei Wang, Chaoping Xie, Xuming Wen, and Qien Yu.
\newblock Upst-nerf: Universal photorealistic style transfer of neural radiance fields for 3d scene.
\newblock \emph{IEEE Transactions on Visualization and Computer Graphics}, 2024{\natexlab{b}}.

\bibitem[Fan et~al.(2025)Fan, Wang, Wen, Zhu, Xu, Wang, et~al.]{fan2025lightgaussian}
Zhiwen Fan, Kevin Wang, Kairun Wen, Zehao Zhu, Dejia Xu, Zhangyang Wang, et~al.
\newblock Lightgaussian: Unbounded 3d gaussian compression with 15x reduction and 200+ fps.
\newblock \emph{Advances in neural information processing systems}, 37:\penalty0 140138--140158, 2025.

\bibitem[Fournier and Reeves(1986)]{fournier1986simple}
Alain Fournier and William~T Reeves.
\newblock A simple model of ocean waves.
\newblock In \emph{Proceedings of the 13th annual conference on Computer graphics and interactive techniques}, pages 75--84, 1986.

\bibitem[Fridovich-Keil et~al.(2023)Fridovich-Keil, Meanti, Warburg, Recht, and Kanazawa]{fridovich2023k}
Sara Fridovich-Keil, Giacomo Meanti, Frederik~Rahb{\ae}k Warburg, Benjamin Recht, and Angjoo Kanazawa.
\newblock K-planes: Explicit radiance fields in space, time, and appearance.
\newblock In \emph{Proceedings of the IEEE/CVF Conference on Computer Vision and Pattern Recognition}, pages 12479--12488, 2023.

\bibitem[Gao et~al.(2024)Gao, Gu, Lin, Li, Zhu, Cao, Zhang, and Yao]{gao2024relightable}
Jian Gao, Chun Gu, Youtian Lin, Zhihao Li, Hao Zhu, Xun Cao, Li Zhang, and Yao Yao.
\newblock Relightable 3d gaussians: Realistic point cloud relighting with brdf decomposition and ray tracing.
\newblock In \emph{European Conference on Computer Vision}, pages 73--89. Springer, 2024.

\bibitem[Green(2004)]{green2004real}
Simon Green.
\newblock Real-time approximations to subsurface scattering.
\newblock \emph{GPU Gems}, 1\penalty0 (263-278):\penalty0 4, 2004.

\bibitem[Guo et~al.(2024)Guo, Ma, Fan, Liu, and Li]{guo2024semantic}
Jun Guo, Xiaojian Ma, Yue Fan, Huaping Liu, and Qing Li.
\newblock Semantic gaussians: Open-vocabulary scene understanding with 3d gaussian splatting.
\newblock \emph{arXiv preprint arXiv:2403.15624}, 2024.

\bibitem[Haque et~al.(2023)Haque, Tancik, Efros, Holynski, and Kanazawa]{haque2023instruct}
Ayaan Haque, Matthew Tancik, Alexei~A Efros, Aleksander Holynski, and Angjoo Kanazawa.
\newblock Instruct-nerf2nerf: Editing 3d scenes with instructions.
\newblock In \emph{Proceedings of the IEEE/CVF International Conference on Computer Vision}, pages 19740--19750, 2023.

\bibitem[Huang et~al.(2022{\natexlab{a}})Huang, Zhang, Feng, Li, Wang, and Wang]{huang2022hdr}
Xin Huang, Qi Zhang, Ying Feng, Hongdong Li, Xuan Wang, and Qing Wang.
\newblock Hdr-nerf: High dynamic range neural radiance fields.
\newblock In \emph{Proceedings of the IEEE/CVF Conference on Computer Vision and Pattern Recognition}, pages 18398--18408, 2022{\natexlab{a}}.

\bibitem[Huang et~al.(2022{\natexlab{b}})Huang, He, Yuan, Lai, and Gao]{huang2022stylizednerf}
Yi-Hua Huang, Yue He, Yu-Jie Yuan, Yu-Kun Lai, and Lin Gao.
\newblock Stylizednerf: consistent 3d scene stylization as stylized nerf via 2d-3d mutual learning.
\newblock In \emph{Proceedings of the IEEE/CVF Conference on Computer Vision and Pattern Recognition}, pages 18342--18352, 2022{\natexlab{b}}.

\bibitem[Ju et~al.(2024)Ju, Liu, Wang, Bian, Shan, and Xu]{ju2024brushnet}
Xuan Ju, Xian Liu, Xintao Wang, Yuxuan Bian, Ying Shan, and Qiang Xu.
\newblock Brushnet: A plug-and-play image inpainting model with decomposed dual-branch diffusion.
\newblock In \emph{European Conference on Computer Vision}, pages 150--168. Springer, 2024.

\bibitem[Ke et~al.(2023)Ke, Liu, Zhu, Zhao, and Lau]{ke2023neural}
Zhanghan Ke, Yuhao Liu, Lei Zhu, Nanxuan Zhao, and Rynson~WH Lau.
\newblock Neural preset for color style transfer.
\newblock In \emph{Proceedings of the IEEE/CVF Conference on Computer Vision and Pattern Recognition}, pages 14173--14182, 2023.

\bibitem[Kerbl et~al.(2023)Kerbl, Kopanas, Leimk{\"u}hler, and Drettakis]{kerbl20233d}
Bernhard Kerbl, Georgios Kopanas, Thomas Leimk{\"u}hler, and George Drettakis.
\newblock 3d gaussian splatting for real-time radiance field rendering.
\newblock \emph{ACM Trans. Graph.}, 42\penalty0 (4):\penalty0 139--1, 2023.

\bibitem[Kerbl et~al.(2024)Kerbl, Meuleman, Kopanas, Wimmer, Lanvin, and Drettakis]{kerbl2024hierarchical}
Bernhard Kerbl, Andreas Meuleman, Georgios Kopanas, Michael Wimmer, Alexandre Lanvin, and George Drettakis.
\newblock A hierarchical 3d gaussian representation for real-time rendering of very large datasets.
\newblock \emph{ACM Transactions on Graphics (TOG)}, 43\penalty0 (4):\penalty0 1--15, 2024.

\bibitem[Kim et~al.(2024)Kim, Youwang, and Oh]{kim2024fprf}
GeonU Kim, Kim Youwang, and Tae-Hyun Oh.
\newblock Fprf: Feed-forward photorealistic style transfer of large-scale 3d neural radiance fields.
\newblock In \emph{Proceedings of the AAAI Conference on Artificial Intelligence}, pages 2750--2758, 2024.

\bibitem[Kingma(2014)]{kingma2014adam}
Diederik~P Kingma.
\newblock Adam: A method for stochastic optimization.
\newblock \emph{arXiv preprint arXiv:1412.6980}, 2014.

\bibitem[Knapitsch et~al.(2017)Knapitsch, Park, Zhou, and Koltun]{Knapitsch2017}
Arno Knapitsch, Jaesik Park, Qian-Yi Zhou, and Vladlen Koltun.
\newblock Tanks and temples: Benchmarking large-scale scene reconstruction.
\newblock \emph{ACM Transactions on Graphics}, 36\penalty0 (4), 2017.

\bibitem[Lenoble et~al.(1985)]{lenoble1985radiative}
Jacqueline Lenoble et~al.
\newblock \emph{Radiative transfer in scattering and absorbing atmospheres: standard computational procedures}.
\newblock A. Deepak Hampton, Va., 1985.

\bibitem[Li and Pan(2024)]{li2024instant}
Shaoxu Li and Ye Pan.
\newblock Instant photorealistic neural radiance fields stylization.
\newblock In \emph{ICASSP 2024-2024 IEEE International Conference on Acoustics, Speech and Signal Processing (ICASSP)}, pages 2980--2984. IEEE, 2024.

\bibitem[Li et~al.(2019)Li, Liu, Kautz, and Yang]{li2019learning}
Xueting Li, Sifei Liu, Jan Kautz, and Ming-Hsuan Yang.
\newblock Learning linear transformations for fast image and video style transfer.
\newblock In \emph{Proceedings of the IEEE/CVF Conference on Computer Vision and Pattern Recognition}, pages 3809--3817, 2019.

\bibitem[Li et~al.(2021)Li, Kou, and Zhao]{li2021weather}
Xuelong Li, Kai Kou, and Bin Zhao.
\newblock Weather gan: Multi-domain weather translation using generative adversarial networks.
\newblock \emph{arXiv preprint arXiv:2103.05422}, 2021.

\bibitem[Li et~al.(2018)Li, Liu, Li, Yang, and Kautz]{li2018closed}
Yijun Li, Ming-Yu Liu, Xueting Li, Ming-Hsuan Yang, and Jan Kautz.
\newblock A closed-form solution to photorealistic image stylization.
\newblock In \emph{Proceedings of the European conference on computer vision (ECCV)}, pages 453--468, 2018.

\bibitem[Li et~al.(2023)Li, Lin, Forsyth, Huang, and Wang]{li2023climatenerf}
Yuan Li, Zhi-Hao Lin, David Forsyth, Jia-Bin Huang, and Shenlong Wang.
\newblock Climatenerf: Extreme weather synthesis in neural radiance field.
\newblock In \emph{Proceedings of the IEEE/CVF International Conference on Computer Vision}, pages 3227--3238, 2023.

\bibitem[Lin et~al.(2014)Lin, Maire, Belongie, Hays, Perona, Ramanan, Doll{\'a}r, and Zitnick]{lin2014microsoft}
Tsung-Yi Lin, Michael Maire, Serge Belongie, James Hays, Pietro Perona, Deva Ramanan, Piotr Doll{\'a}r, and C~Lawrence Zitnick.
\newblock Microsoft coco: Common objects in context.
\newblock In \emph{Computer Vision--ECCV 2014: 13th European Conference, Zurich, Switzerland, September 6-12, 2014, Proceedings, Part V 13}, pages 740--755. Springer, 2014.

\bibitem[Liu et~al.(2023)Liu, Zhan, Chen, Zhang, Yu, El~Saddik, Lu, and Xing]{liu2023stylerf}
Kunhao Liu, Fangneng Zhan, Yiwen Chen, Jiahui Zhang, Yingchen Yu, Abdulmotaleb El~Saddik, Shijian Lu, and Eric~P Xing.
\newblock Stylerf: Zero-shot 3d style transfer of neural radiance fields.
\newblock In \emph{Proceedings of the IEEE/CVF Conference on Computer Vision and Pattern Recognition}, pages 8338--8348, 2023.

\bibitem[Liu et~al.(2024{\natexlab{a}})Liu, Zhan, Xu, Theobalt, Shao, and Lu]{liu2024stylegaussian}
Kunhao Liu, Fangneng Zhan, Muyu Xu, Christian Theobalt, Ling Shao, and Shijian Lu.
\newblock Stylegaussian: Instant 3d style transfer with gaussian splatting.
\newblock In \emph{SIGGRAPH Asia 2024 Technical Communications}, pages 1--4. 2024{\natexlab{a}}.

\bibitem[Liu et~al.(2024{\natexlab{b}})Liu, Xue, Luo, Tan, and Yi]{liu2024genn2n}
Xiangyue Liu, Han Xue, Kunming Luo, Ping Tan, and Li Yi.
\newblock Genn2n: Generative nerf2nerf translation.
\newblock In \emph{Proceedings of the IEEE/CVF Conference on Computer Vision and Pattern Recognition}, pages 5105--5114, 2024{\natexlab{b}}.

\bibitem[Mei et~al.(2025)Mei, Xu, and Patel]{mei2025regs}
Yiqun Mei, Jiacong Xu, and Vishal Patel.
\newblock Regs: Reference-based controllable scene stylization with gaussian splatting.
\newblock \emph{Advances in Neural Information Processing Systems}, 37:\penalty0 4035--4061, 2025.

\bibitem[Miao et~al.(2024)Miao, Bai, Duan, Wan, Huang, Long, and Zheng]{miao2024conrf}
Xingyu Miao, Yang Bai, Haoran Duan, Fan Wan, Yawen Huang, Yang Long, and Yefeng Zheng.
\newblock Conrf: Zero-shot stylization of 3d scenes with conditioned radiation fields.
\newblock \emph{arXiv preprint arXiv:2402.01950}, 2024.

\bibitem[M{\"u}ller et~al.(2022)M{\"u}ller, Evans, Schied, and Keller]{muller2022instant}
Thomas M{\"u}ller, Alex Evans, Christoph Schied, and Alexander Keller.
\newblock Instant neural graphics primitives with a multiresolution hash encoding.
\newblock \emph{ACM transactions on graphics (TOG)}, 41\penalty0 (4):\penalty0 1--15, 2022.

\bibitem[Nishita et~al.(1987)Nishita, Miyawaki, and Nakamae]{nishita1987shading}
Tomoyuki Nishita, Yasuhiro Miyawaki, and Eihachiro Nakamae.
\newblock A shading model for atmospheric scattering considering luminous intensity distribution of light sources.
\newblock \emph{Acm Siggraph Computer Graphics}, 21\penalty0 (4):\penalty0 303--310, 1987.

\bibitem[Nishita et~al.(1997)Nishita, Iwasaki, Dobashi, and Nakamae]{nishita1997modeling}
Tomoyuki Nishita, Hiroshi Iwasaki, Yoshinori Dobashi, and Eihachiro Nakamae.
\newblock A modeling and rendering method for snow by using metaballs.
\newblock In \emph{Computer Graphics Forum}, pages C357--C364. Wiley Online Library, 1997.

\bibitem[Oh et~al.(2024)Oh, Kim, Kim, and Kim]{oh2024monowad}
Youngmin Oh, Hyung-Il Kim, Seong~Tae Kim, and Jung~Uk Kim.
\newblock Monowad: Weather-adaptive diffusion model for robust monocular 3d object detection.
\newblock In \emph{European Conference on Computer Vision}, pages 326--345. Springer, 2024.

\bibitem[Sakaridis et~al.(2018)Sakaridis, Dai, and Van~Gool]{sakaridis2018semantic}
Christos Sakaridis, Dengxin Dai, and Luc Van~Gool.
\newblock Semantic foggy scene understanding with synthetic data.
\newblock \emph{International Journal of Computer Vision}, 126:\penalty0 973--992, 2018.

\bibitem[Schlick(1994)]{schlick1994inexpensive}
Christophe Schlick.
\newblock An inexpensive brdf model for physically-based rendering.
\newblock In \emph{Computer graphics forum}, pages 233--246. Wiley Online Library, 1994.

\bibitem[Schmidt et~al.(2022)Schmidt, Luccioni, Teng, Zhang, Reynaud, Raghupathi, Cosne, Juraver, Vardanyan, Hern{\'a}ndez-Garc{\'\i}a, and Bengio]{schmidt2022climategan}
Victor Schmidt, Alexandra Luccioni, M{\'e}lisande Teng, Tianyu Zhang, Alexia Reynaud, Sunand Raghupathi, Gautier Cosne, Adrien Juraver, Vahe Vardanyan, Alex Hern{\'a}ndez-Garc{\'\i}a, and Yoshua Bengio.
\newblock Climate{GAN}: Raising climate change awareness by generating images of floods.
\newblock In \emph{International Conference on Learning Representations}, 2022.

\bibitem[Stomakhin et~al.(2013)Stomakhin, Schroeder, Chai, Teran, and Selle]{stomakhin2013material}
Alexey Stomakhin, Craig Schroeder, Lawrence Chai, Joseph Teran, and Andrew Selle.
\newblock A material point method for snow simulation.
\newblock \emph{ACM Transactions on Graphics (TOG)}, 32\penalty0 (4):\penalty0 1--10, 2013.

\bibitem[Sun et~al.(2020)Sun, Kretzschmar, Dotiwalla, Chouard, Patnaik, Tsui, Guo, Zhou, Chai, Caine, Vasudevan, Han, Ngiam, Zhao, Timofeev, Ettinger, Krivokon, Gao, Joshi, Zhang, Shlens, Chen, and Anguelov]{Sun_2020_CVPR}
Pei Sun, Henrik Kretzschmar, Xerxes Dotiwalla, Aurelien Chouard, Vijaysai Patnaik, Paul Tsui, James Guo, Yin Zhou, Yuning Chai, Benjamin Caine, Vijay Vasudevan, Wei Han, Jiquan Ngiam, Hang Zhao, Aleksei Timofeev, Scott Ettinger, Maxim Krivokon, Amy Gao, Aditya Joshi, Yu Zhang, Jonathon Shlens, Zhifeng Chen, and Dragomir Anguelov.
\newblock Scalability in perception for autonomous driving: Waymo open dataset.
\newblock In \emph{Proceedings of the IEEE/CVF Conference on Computer Vision and Pattern Recognition (CVPR)}, 2020.

\bibitem[Tan and Le(2019)]{tan2019efficientnet}
Mingxing Tan and Quoc Le.
\newblock Efficientnet: Rethinking model scaling for convolutional neural networks.
\newblock In \emph{International conference on machine learning}, pages 6105--6114. PMLR, 2019.

\bibitem[Tessendorf et~al.(2001)]{tessendorf2001simulating}
Jerry Tessendorf et~al.
\newblock Simulating ocean water.
\newblock \emph{Simulating nature: realistic and interactive techniques. SIGGRAPH}, 1\penalty0 (2):\penalty0 5, 2001.

\bibitem[Verbin et~al.(2022)Verbin, Hedman, Mildenhall, Zickler, Barron, and Srinivasan]{verbin2022ref}
Dor Verbin, Peter Hedman, Ben Mildenhall, Todd Zickler, Jonathan~T Barron, and Pratul~P Srinivasan.
\newblock Ref-nerf: Structured view-dependent appearance for neural radiance fields.
\newblock In \emph{2022 IEEE/CVF Conference on Computer Vision and Pattern Recognition (CVPR)}, pages 5481--5490. IEEE, 2022.

\bibitem[Wang et~al.(2022)Wang, Chai, He, Chen, and Liao]{wang2022clip}
Can Wang, Menglei Chai, Mingming He, Dongdong Chen, and Jing Liao.
\newblock Clip-nerf: Text-and-image driven manipulation of neural radiance fields.
\newblock In \emph{Proceedings of the IEEE/CVF Conference on Computer Vision and Pattern Recognition}, pages 3835--3844, 2022.

\bibitem[Wang et~al.(2024)Wang, Fang, Zhang, Xie, and Tian]{wang2024gaussianeditor}
Junjie Wang, Jiemin Fang, Xiaopeng Zhang, Lingxi Xie, and Qi Tian.
\newblock Gaussianeditor: Editing 3d gaussians delicately with text instructions.
\newblock In \emph{Proceedings of the IEEE/CVF Conference on Computer Vision and Pattern Recognition}, pages 20902--20911, 2024.

\bibitem[Wu et~al.(2024)Wu, Bian, Li, Wang, Reid, Torr, and Prisacariu]{wu2024gaussctrl}
Jing Wu, Jia-Wang Bian, Xinghui Li, Guangrun Wang, Ian Reid, Philip Torr, and Victor~Adrian Prisacariu.
\newblock Gaussctrl: Multi-view consistent text-driven 3d gaussian splatting editing.
\newblock In \emph{European Conference on Computer Vision}, pages 55--71. Springer, 2024.

\bibitem[Yan et~al.(2024)Yan, Lin, Zhou, Wang, Sun, Zhan, Lang, Zhou, and Peng]{yan2024street}
Yunzhi Yan, Haotong Lin, Chenxu Zhou, Weijie Wang, Haiyang Sun, Kun Zhan, Xianpeng Lang, Xiaowei Zhou, and Sida Peng.
\newblock Street gaussians: Modeling dynamic urban scenes with gaussian splatting.
\newblock In \emph{European Conference on Computer Vision}, pages 156--173. Springer, 2024.

\bibitem[Yang et~al.(2024)Yang, Kang, Huang, Xu, Feng, and Zhao]{yang2024depth}
Lihe Yang, Bingyi Kang, Zilong Huang, Xiaogang Xu, Jiashi Feng, and Hengshuang Zhao.
\newblock Depth anything: Unleashing the power of large-scale unlabeled data.
\newblock In \emph{Proceedings of the IEEE/CVF Conference on Computer Vision and Pattern Recognition}, pages 10371--10381, 2024.

\bibitem[Ye et~al.(2024)Ye, Danelljan, Yu, and Ke]{ye2024gaussian}
Mingqiao Ye, Martin Danelljan, Fisher Yu, and Lei Ke.
\newblock Gaussian grouping: Segment and edit anything in 3d scenes.
\newblock In \emph{European Conference on Computer Vision}, pages 162--179. Springer, 2024.

\bibitem[Yu et~al.(2024)Yu, Chen, Huang, Sattler, and Geiger]{yu2024mip}
Zehao Yu, Anpei Chen, Binbin Huang, Torsten Sattler, and Andreas Geiger.
\newblock Mip-splatting: Alias-free 3d gaussian splatting.
\newblock In \emph{Proceedings of the IEEE/CVF Conference on Computer Vision and Pattern Recognition}, pages 19447--19456, 2024.

\bibitem[Zhang et~al.(2023{\natexlab{a}})Zhang, Rao, and Agrawala]{zhang2023adding}
Lvmin Zhang, Anyi Rao, and Maneesh Agrawala.
\newblock Adding conditional control to text-to-image diffusion models, 2023{\natexlab{a}}.

\bibitem[Zhang et~al.(2023{\natexlab{b}})Zhang, Liu, Han, Pan, Guo, and Yao]{zhang2023transforming}
Zicheng Zhang, Yinglu Liu, Congying Han, Yingwei Pan, Tiande Guo, and Ting Yao.
\newblock Transforming radiance field with lipschitz network for photorealistic 3d scene stylization.
\newblock In \emph{Proceedings of the IEEE/CVF Conference on Computer Vision and Pattern Recognition}, pages 20712--20721, 2023{\natexlab{b}}.

\bibitem[Zhou et~al.(2024)Zhou, Chang, Jiang, Fan, Zhu, Xu, Chari, You, Wang, and Kadambi]{zhou2024feature}
Shijie Zhou, Haoran Chang, Sicheng Jiang, Zhiwen Fan, Zehao Zhu, Dejia Xu, Pradyumna Chari, Suya You, Zhangyang Wang, and Achuta Kadambi.
\newblock Feature 3dgs: Supercharging 3d gaussian splatting to enable distilled feature fields.
\newblock In \emph{Proceedings of the IEEE/CVF Conference on Computer Vision and Pattern Recognition}, pages 21676--21685, 2024.

\bibitem[Zhu et~al.(2017)Zhu, Park, Isola, and Efros]{zhu2017unpaired}
Jun-Yan Zhu, Taesung Park, Phillip Isola, and Alexei~A Efros.
\newblock Unpaired image-to-image translation using cycle-consistent adversarial networks.
\newblock In \emph{Proceedings of the IEEE international conference on computer vision}, pages 2223--2232, 2017.

\end{thebibliography}
